\documentclass[prb,twocolumn,superscriptaddress]{revtex4-1}
\usepackage{graphicx}
\usepackage{bm}
\usepackage{amssymb}
\usepackage{amsmath}
\usepackage{upgreek}

\begin{document}
\title{Interlayer coupling in commensurate and incommensurate bilayer structures of transition metal dichalcogenides}

\author{Yong Wang}
\affiliation{School of Physics, Nankai University, Tianjin 300071, China}
\affiliation{Department of Physics and Center of Theoretical and Computational Physics, The University of Hong Kong, Hong Kong, China}

\author{Zhan Wang}
\affiliation{School of Physics, Nankai University, Tianjin 300071, China}

\author{Wang Yao}
\affiliation{Department of Physics and Center of Theoretical and Computational Physics, The University of Hong Kong, Hong Kong, China}

\author{Gui-Bin Liu}
\email[]{gbliu@bit.edu.cn}
\affiliation{School of Physics, Beijing Institute of Technology, Beijing 100081, China}

\author{Hongyi Yu}
\email[]{yuhongyi@hku.hk}
\affiliation{Department of Physics and Center of Theoretical and Computational Physics, The University of Hong Kong, Hong Kong, China}

\begin{abstract}
The interlayer couplings in commensurate and incommensurate bilayer structures of transition metal dichalcogenides are investigated with perturbative treatment. The interlayer coupling in $\pm\mathbf K$ valleys can be decomposed into a series of hopping terms with distinct phase factors. In H-type and R-type commensurate bilayers, the interference between the three main hopping terms leads to a sensitive dependence of the interlayer coupling strength on the translation, that can explain the position dependent local band gap modulation in a heterobilayer moir\'{e} superlattice. The interlayer couplings in the $\bm\Gamma$ valley of valence band and $\mathbf Q$ valley of conduction band are also studied, where the strong coupling strengths of several hundred meV can play important roles in mediating the ultrafast interlayer charge transfer in heterobilayers of transition metal dichalcogenides.
\end{abstract}

\maketitle

\section{Introduction}
Monolayer group-VIB transition metal dichalcogenides (TMDs) have been extensively studied in recent years, mainly due to their exotic physical properties and potential applications in novel two-dimensional (2D) electronics devices.\cite{Schaibley2016NRM,Review_kpTheory,Mak2016Review,Jariwala2014ACSnano,Liu2015CSR} Compared with the precedent 2D material graphene, monolayer TMDs have a finite and direct band gap located at the two degenerate but inequivalent hexagonal Brillouin zone (BZ) corners, i.e., the $\pm\mathbf K$ valleys, which are essential for the successful operation of transistors and valley-dependent optoelectronics. Furthermore, the strong spin-orbit coupling of the transition metal couples the spin and valley degrees of freedom, making TMDs the ideal platform to develop spintronic and valleytronic devices.\cite{Schaibley2016NRM} Several kinds of electronic and optoelectronic prototype devices have been fabricated with monolayer TMDs, including field-effect transistor, inverter and logic gate, junction and heterostructure, photodetector, solar cell and light-emitting devices, as well as electronic sensors.\cite{Mak2016Review,Jariwala2014ACSnano}

Similar to the monolayers, the natural TMD homobilayers can be obtained from bulk crystals using mechanical exfoliation and have been widely studied. These natural homobilayers mostly exhibit a commensurate 2H (also called AB) stacking where the two layers are $180^\circ$ rotation of each other.\cite{Liu2015CSR} As the two adjacent layers are bound together by the weak van der Waals interaction, the interlayer coupling in $\pm\mathbf K$ valleys can be largely suppressed by the giant spin-orbit splitting. The resulted spin-layer locking could lead to various magnetoelectric effects allowing for their quantum manipulations. \cite{Gong2013NC,Wu2013NP,Jones2014NP,Yuan2013NP,Zhu2014PNAS,Jiang2014NN} On the other hand, the interlayer couplings in the valence band $\bm\Gamma$ and conduction band $\mathbf Q$ valleys are significantly larger, which strongly shifts their energy positions compared to those of the monolayers and results in a transition from direct to indirect band gap.\cite{Mak2010PRL,Bradley2015NL}

Furthermore, the current technique allows manually stacking two monolayers to form a vertical homo- or heterostructure, with the uncertainty lesser than  $1^\circ$ on their mutual crystallographic alignment.\cite{Ponomarenko2013Nature,Dean2013Nature,Hunt2013Science} This opens up an alternative way to utilize this novel class of 2D materials.\cite{Geim2013Nature} For the TMD heterobilayer formed by two different TMD materials, its conduction and valence band edges are located in different layers. Such a type-II band alignment results in the ultrafast interlayer charge transfer which facilitates the photocurrent generation,\cite{Fang_PNAS,Chiu_ACSNano,Lee_NatNano,Furchi_NL,Cheng_NL,Ceballos_2014ACSNano,Hong_NatNano,Yu_2015NL,Rigosi_2015NL} and the formation of interlayer excitons.\cite{Rivera_InterlayerX0,Yu2015PRL,Rivera2016Science} Meanwhile, the manually assembled bilayer generally has an incommensurate lattice structure due to the inevitable interlayer twist and/or lattice constant mismatch. This brings anomalous interlayer couplings which have profound effects on the transport, \cite{Ponomarenko2013Nature,Dean2013Nature,Hunt2013Science,Jung2015NC,Zhou2016arXiv} optical \cite{Liu2014NC,vanderZande2014NL,Hsu2014ACSnano,Heo2015NC,Huang2014NL} and Raman \cite{Puretzky2016ACSnano,Lui2015PRB} properties of the bilayers. Moreover, recent theoretical studies have shown that  the interlayer coupling together with the formation of a large scale moir\'{e} superlattice pattern can lead to the emergence of topological orders in a TMD heterobilayer. \cite{Tong2016arXiv,MacDonald2016arXiv} To gain further insights into these interesting phenomena, it is essential to understand the strength and the form of the interlayer coupling in TMD bilayers.

In 2H or other commensurate bilayers, the interlayer coupling can be evaluated by comparing the bilayer band structure to those of the monolayers. The $2\pi/3$-rotational symmetry of the 2H bilayer is also essential to determine whether the interlayer coupling strength at $\pm\mathbf K$ points is zero or not.\cite{Liu2015CSR} For the general TMD bilayers, however, it is non-trivial to calculate the interlayer coupling of the incommensurate lattice structures mainly due to the lack of periodic feature. For the limited commensurate cases, the unit cell usually contains too many atoms to be calculated from first principles. Thus some analytical way should be adopted instead of the impractical numerical calculations.

In this paper, we investigate the interlayer coupling in general TMD bilayers following the previous studies in twisted bilayer graphene,\cite{Santos2007PRL,Bistritzer2011PNAS,Shallcross2010PRB,Santos2012PRB,SanJose2014PRB} by adopting an effective perturbative treatment. The rest of the paper is organized as follows. In section II we show that, in general TMD bilayers the interlayer coupling between the $\pm\mathbf{K}$ valley Bloch states can be decomposed into a series of hopping terms with distinct phase factors, which correspond to the Fourier components of the hopping integral between localized atomic orbitals. In section III, the symmetry properties of the monolayer TMDs are analyzed and utilized to reveal the relation between the hopping terms. In section IV, our perturbative results for the commensurate H- and R-type TMD homobilayers are presented, which show sensitive dependence on the interlayer translation, and are in excellent agreement with the \textit{ab initio} calculations. In section V, we apply our perturbative treatment to the lattice-mismatched bilayers, and reveal its connection with the moir\'{e} superlattice. In section VI, we further study the interlayer coupling of the valence band $\mathbf \Gamma$ and conduction band $\mathbf Q$ valleys, and propose that they play important roles in mediating the ultrafast interlayer charge transfer of TMD heterobilayers. We summarize our results in section VII.

\section{Expression of interlayer coupling in $\pm\mathbf K$ valleys}

Since the two TMD monolayers are bound by the weak van der Waals force, we can first consider a decoupled bilayer, then add the interlayer coupling as a perturbation. In the vanishing interlayer coupling limit, the monolayer Bloch wavefunctions in $\tau\mathbf{K}$ valley are denoted as
$\psi_{n,\mathbf k}(\mathbf r)\equiv\langle\mathbf r|n,\mathbf k\rangle=e^{i(\tau\mathbf{K}+\mathbf{k})\cdot\mathbf{r}}u_{n,\mathbf k}(\mathbf r)$. Here, $n=\{\tau,l\}$ contains both the valley index $\tau=\pm$ and the band index $l=\cdots,c+1,c,v,v-1,\cdots$. Here $c$ ($v$) corresponds to the conduction (valence) band, and we use $c+j$
$(v-j)$ to denote the $j$-th band above (below) the conduction (valence) band. $u_{n,\mathbf k}(\mathbf r)$ is the periodic part of the Bloch wavefunctions.

The Bloch wavefunction $\psi_{n,0}$ can be constructed from the local basis functions as
\begin{align}
\psi_{n,0}(\mathbf{r})=\frac{1}{\sqrt{N}}\sum_{\mathbf R}e^{i\tau\mathbf{K}\cdot\mathbf{R}}D_n(\mathbf{r}-\mathbf{R}).
\label{Bloch1}
\end{align}
Here, $N$ is the unit cell number of the corresponding monolayer, $D_n(\mathbf{r}-\mathbf{R})$ is the linear combination of the atomic orbitals localized near the metal position $\mathbf{R}$, which depends on the valley index $\tau$ and band index $l$ (see Table \ref{bands_C3_quantum_number}). Considering the time reversal relation between the two valleys, $D_n$ in the same band but opposite valleys are related by a complex conjugate. Under the envelope approximation, $\psi_{n,\mathbf k}(\mathbf r)\approx e^{i(\tau\mathbf K+\mathbf k)\cdot\mathbf r}u_{n,0}(\mathbf r)=e^{i\mathbf k\cdot\mathbf r}\psi_{n,0}(\mathbf r)$, one finds
\begin{align}
\psi_{n,\mathbf k}(\mathbf r)\approx\frac{1}{\sqrt{N}}\sum_{\mathbf R}e^{i(\tau\mathbf K+\mathbf k)\cdot\mathbf R}D_n(\mathbf r-\mathbf R),
\label{Bloch2}
\end{align}
where $e^{i\mathbf k\cdot(\mathbf r-\mathbf R)}D_n(\mathbf r-\mathbf R)\approx D_n(\mathbf r-\mathbf R)$ is used since we are interested in low energy electrons and holes with small $|\mathbf k|$, and $D_n(\mathbf r-\mathbf R)$ is well localized near $\mathbf R$.

\begin{table}[htbp]
\begin{center}
\caption{The orbital compositions and the corresponding $2\pi/3$-rotational quantum numbers $M$ (discussed in Section III A) of localized function $D_n$ in $+\mathbf K$ valley of monolayer MoS$_2$, obtained by following Ref. [\onlinecite{Liu2015CSR}]. The percentage is defined as the overlap probability between the atomic orbital wave function and the $\mathbf K$-point Bloch state. $d_0\equiv d_{z^2}$, $d_{\pm1}\equiv(d_{xz}\pm id_{yz})/\sqrt{2}$, $d_{\pm2}\equiv(d_{x^2-y^2}\pm id_{xy})/\sqrt{2}$ are the Mo-$d$ orbitals, and $p_0\equiv p_z$, $p_{\pm1}\equiv(p_x\pm ip_y)/\sqrt{2}$ are the S-$p$ orbitals. Mo-$p_0$ denotes the $p_z$ orbital of the Mo atom. Only the two most prominent orbitals are shown. }
\label{bands_C3_quantum_number}
\begin{tabular}{cccc}
\hline\hline
(Band) & (Major orbital) & (Minor orbital) & ($M$) \\
\hline 
$\vdots$ & $\vdots$ & $\vdots$ & $\vdots$ \\
$c+3$ & $d_{+1}$ (70\%) & $p_0$ (24\%) & $+1$ \\
$c+2$ & $d_{-2}$ (78\%) & $p_0$ (19\%) & $+1$ \\
$c+1$ & $d_{-1}$ (78\%) & $p_{+1}$ (22\%) & $-1$ \\
$c$ & $d_0$ (88\%) & $p_{-1}$ (7\%) & $0$ \\
$v$ & $d_{+2}$ (84\%) & $p_{+1}$ (16\%) & $-1$ \\
$v-1$ & $p_0$ (56\%) & $d_{+1}$ (38\%) & $+1$ \\
$v-2$ & $p_{-1}$ (83\%) & Mo-$p_0$ (17\%) & $0$ \\
$v-3$ & $p_0$ (53\%) & $d_{-2}$ (31\%) & $+1$ \\
$\vdots$ & $\vdots$ & $\vdots$ & $\vdots$ \\
\hline\hline
\end{tabular}
\end{center}
\end{table}

\begin{figure}[]
\includegraphics[width=\linewidth]{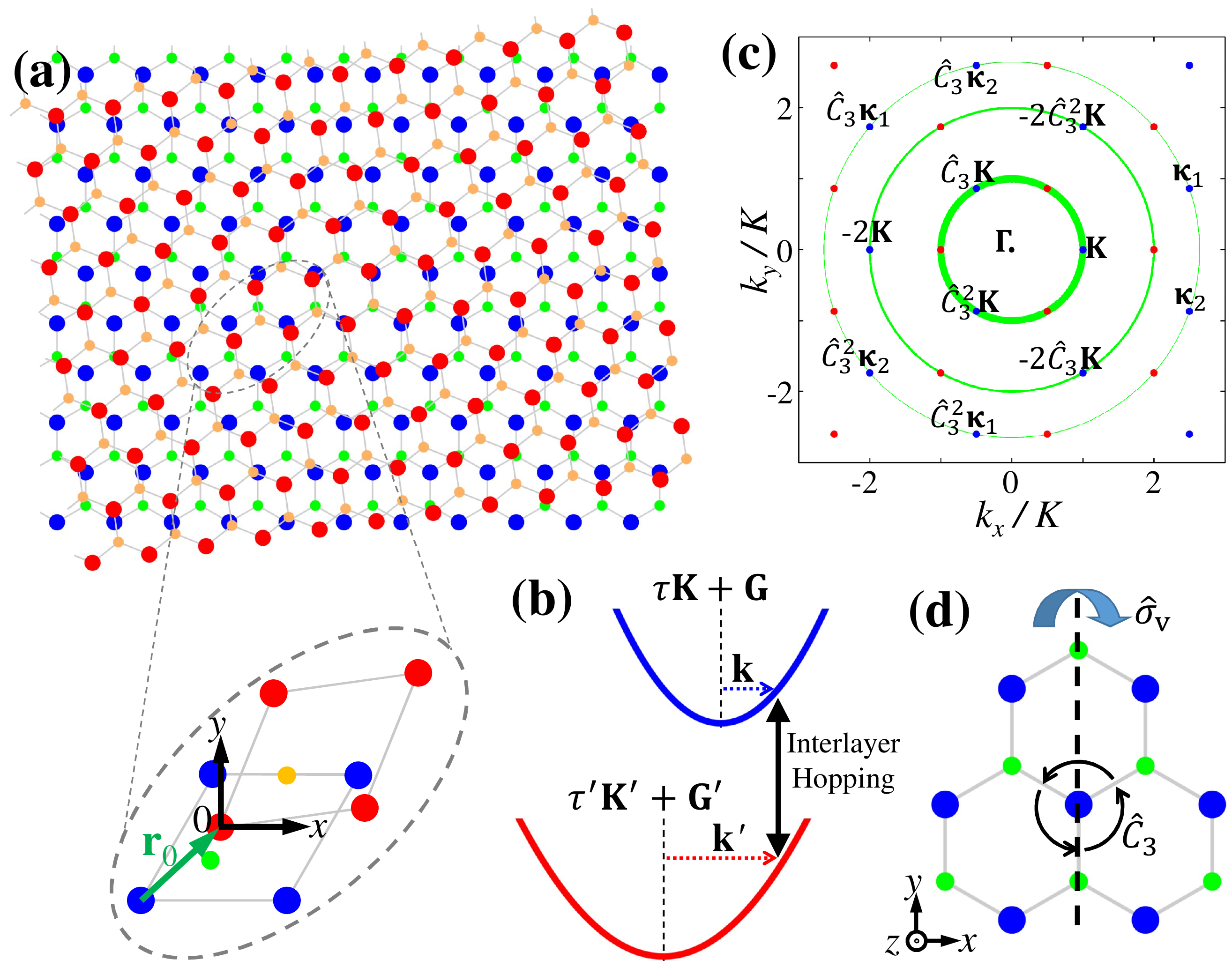}
\caption{(Color online) (a) Illustration of a twisted TMD homo- or heterobilayer. The large red (blue) dots denote the metal atoms in the upper (lower) layer, and the small orange (green) dots denote the chalcogen atoms in the upper (lower) layer. The enlarged view shows two unit cells in the upper and lower layers, respectively. The in-plane ($xy$) coordinate origin is set on a metal atom in the upper layer. (b) Two wave vectors in different layers must overlap in momentum space to satisfy the momentum conservation of interlayer hopping. (c) The blue dots denote $\bm\upkappa\equiv\mathbf K+\mathbf G$ points, and the red dots are their time reversals. Thicker green circle means smaller $|\bm\upkappa|$ thus larger $|t_{nn'}(\tau\bm\upkappa)|$ (see Eq. (\ref{Hopping3})). (d) Illustration of the $2\pi/3$-rotational ($\hat C_3$) symmetry and the in-plane mirror ($\hat\sigma_\mathrm{v}$) symmetry of monolayer TMDs.}
\label{Fig1}
\end{figure}

We define a bilayer stacking configuration as the reference one, where the in-plane crystalline axes of the two layers are along the same direction (R-type stacking), and the two metal atoms in different layers horizontally overlap at the in-plane ($xy$) coordinate origin. Any other stacking configuration can then be obtained from this reference configuration through a $\theta$-angle rotation of the upper layer around the coordinate origin, and followed by a translation of $-\mathbf r_0$ for the lower layer (see Fig.~\ref{Fig1}(a)). We use the convention that quantities in the upper (lower) layer are marked with (without) the prime. The lower layer band edges are located at $\pm\mathbf K=\pm\frac{4\pi}{3a}(1,0)$, while those of the upper layer are located at $\pm\mathbf K'=\pm\frac{4\pi}{3a'}(\cos\theta,\sin\theta)$, where $a$ ($a'$) is the lower (upper) layer lattice constant.

Now we add the interlayer coupling $\hat H_t$ as a perturbation. We consider the hopping integral between the two wavefunctions $\psi_{n',\mathbf k'}$ and $\psi_{n,\mathbf k}$ located in the upper and lower layer respectively, which can be expressed as
\begin{align}
&\langle n,\mathbf k|\hat H_t|n',\mathbf k'\rangle\equiv\int\psi^*_{n,\mathbf k}(\mathbf r)\hat H_t\psi_{n',\mathbf k'}(\mathbf r)d\mathbf r\nonumber\\
=&\sum_{\mathbf R,\mathbf R'}\frac{e^{i(\tau'\mathbf K'+\mathbf k')\cdot\mathbf R'-i(\tau\mathbf K+\mathbf k)\cdot\mathbf R}}{\sqrt{NN'}}\langle D_{n,\mathbf R}|\hat H_t|D_{n',\mathbf R'}\rangle.
\label{Hopping1}
\end{align}
Here, $\langle D_{n,\mathbf R}|\hat H_t|D_{n',\mathbf R'}\rangle\equiv\int D^*_n(\mathbf r-\mathbf R)\hat H_tD_{n'}(\mathbf r-\mathbf R')d\mathbf r$ is the hopping integral between the two localized orbitals
$D_{n'}(\mathbf r-\mathbf R')$ and $D_n(\mathbf r-\mathbf R)$. In the spirit of two-center approximation,\cite{Santos2007PRL,Bistritzer2011PNAS,Shallcross2010PRB} $\langle D_{n,\mathbf R}|\hat H_t|D_{n',\mathbf R'}\rangle$ depends only on the relative position $\mathbf R'-\mathbf R$. So we can write
\begin{align}
\langle D_{n,\mathbf R}|\hat H_t|D_{n',\mathbf R'}\rangle&=T_{nn'}(\mathbf R'-\mathbf R)\nonumber\\
&=\sum_{\mathbf q}\frac{e^{-i\mathbf q\cdot(\mathbf R'-\mathbf R)}}{\sqrt{NN'}}t_{nn'}(\mathbf q).
\label{HoppingFourierTransf}
\end{align}
Here, $t_{nn'}(\mathbf q)=\frac{1}{\sqrt{\Omega\Omega'}}\int T_{nn'}(\mathbf r)e^{i\mathbf q\cdot\mathbf r}d\mathbf r$ is the Fourier transform of $T_{nn'}(\mathbf r)$, with $\Omega'$ ($\Omega$) the upper (lower) layer unit cell area.

We denote the in-plane positions of the the metal atoms in the upper (lower) layer as $\mathbf R'=j'_1\mathbf a'_1+j'_2\mathbf a'_2$ ($\mathbf R=-\mathbf r_0+j_1\mathbf a_1+j_2\mathbf a_2$), where $\mathbf a'_{1,2}$ ($\mathbf a_{1,2}$) are the corresponding unit lattice vectors and $j'_{1,2},j_{1,2}$ are integers. Substituting Eq. (\ref{HoppingFourierTransf}) into Eq. (\ref{Hopping1}), we obtain
\begin{align}
&\langle n,\mathbf k|\hat H_t|n',\mathbf k'\rangle \nonumber\\
=&\sum_{\mathbf q}t_{nn'}(\mathbf q)
\sum_{\mathbf R,\mathbf R'}\frac{e^{i(\tau'\mathbf K'+\mathbf k'-\mathbf q)\cdot\mathbf R'-i(\tau\mathbf K+\mathbf k-\mathbf q)\cdot\mathbf R}}{NN'}\nonumber\\
=&\sum_{\mathbf q}t_{nn'}(\mathbf q)\sum_{\mathbf G,\mathbf G'}\delta_{\tau\mathbf K+\mathbf k-\mathbf q,\mathbf G}\delta_{\tau'\mathbf K'+\mathbf k'-\mathbf q,\mathbf G'}e^{i\mathbf G\cdot\mathbf r_0}\nonumber\\
=&\sum_{\mathbf G,\mathbf G'}\delta_{\tau\mathbf K+\mathbf k+\mathbf G,\tau'\mathbf K'+\mathbf k'+\mathbf G'}t_{nn'}(\tau\mathbf K+\mathbf k+\mathbf G)e^{-i\mathbf G\cdot\mathbf r_0}.  \nonumber
\end{align}
Here, $\mathbf G'$ ($\mathbf G)$ is the reciprocal lattice vector of the upper (lower) layer. Note that the translation vector $\mathbf r_0$ appear in the phase factor $e^{-i\mathbf G\cdot\mathbf r_0}$ only. Since the phase factor doesn't change when we replace $\mathbf r_0$ by $\mathbf r_0+j_1\mathbf a_1+j_2\mathbf a_2$, we can restrict $\mathbf r_0$ to be inside a unit cell of the lower layer.

To simplify the above expression, we use the notation $\tau'\bm\upkappa'\equiv\tau'\mathbf K'+\mathbf G'$ and $\tau\bm\upkappa\equiv\tau\mathbf K+\mathbf G$, and write
\begin{align}
&\langle n,\mathbf k|\hat H_t|n',\mathbf k'\rangle \nonumber\\
=~&e^{i\tau\mathbf K\cdot\mathbf r_0}\sum_{\bm\upkappa'\bm\upkappa}\delta_{\mathbf k'-\mathbf k,\tau\bm\upkappa-\tau'\bm\upkappa'}t_{nn'}(\tau\bm\upkappa+\mathbf{k})e^{-i\tau\bm\upkappa\cdot\mathbf r_0}.
\label{Hopping3}
\end{align}
Eq.~(\ref{Hopping3}) is the central result of this paper, closely similar forms also appear in other works for graphene-related van der Waals materials \cite{Santos2007PRL,Bistritzer2011PNAS,Shallcross2010PRB,Santos2012PRB,SanJose2014PRB} and our early paper for heterobilayer TMDs.\cite{Tong2016arXiv} It implies that the hopping integral between two Bloch functions in different layers is nonzero only when $\mathbf k'-\mathbf k$ equals one of the discrete values $\tau\bm\upkappa-\tau'\bm\upkappa'$, as illustrated in Fig. \ref{Fig1}(b). Furthermore we expect $t_{nn'}(\mathbf q)$ to decay fast with the increase of $|\mathbf q|$, as $D_n(\mathbf r)$ and $D_{n'}(\mathbf r)$ vary smoothly with $\mathbf r$ and the integral $\langle D_{n,\mathbf R}|\hat H_t|D_{n',\mathbf R'}\rangle$ is generally a smooth function of $\mathbf R'-\mathbf R$. Therefore, in the summation $\sum_{\bm\upkappa'\bm\upkappa}$ only a few terms of $\bm\upkappa'$ and $\bm\upkappa$ with small magnitudes need to be kept, which greatly reduces the number of $\tau\bm\upkappa-\tau'\bm\upkappa'$. In Fig.~\ref{Fig1}(c), we show three groups of $\bm\upkappa$. $\mathbf K$, $\hat{C}_3\mathbf K$ and $\hat{C}^2_3\mathbf K$ on the thickest circle are closest to $\bm\Gamma$ and are expected to have the most pronounced $|t_{nn'}|$; $-2\mathbf K$, $-2\hat{C}_3\mathbf K$ and $-2\hat{C}^2_3\mathbf K$ ($\bm\upkappa_{1,2}$, $\hat{C}_3\bm\upkappa_{1,2}$ and $\hat C^2_3\bm\upkappa_{1,2}$) are the second (third) closest to $\bm\Gamma$, and the corresponding $|t_{nn'}|$ values are expected to be much weaker.

\section{Symmetry properties of the hopping terms}
The monolayer hexagonal lattice structure has both the $2\pi/3$-rotational ($\hat{C}_{3}$) symmetry and the in-plane mirror ($\hat\sigma_\mathrm{v}$) symmetry (see Fig. \ref{Fig1}(d)). The hopping terms $t_{nn'}(\mathbf{q})$ with the same $|\mathbf q|$ values but different $\mathbf{q}$ directions are related by these symmetry operations.

\subsection{$2\pi/3$-rotational symmetry}
We use $\hat{C}_3$ to denote the in-plane counter-clockwise $\frac{2\pi}{3}$-rotation around $\mathbf r=0$ when applied on a real space vector (around $\bm\Gamma$-point when applied on a $\mathbf k$-space vector). As the $\tau\mathbf K$ point has a high symmetry, i.e., $\hat{C}_3\tau\mathbf K=\tau\mathbf K+\mathbf G$, the orbital combination $D_n(\mathbf r)$ should be $\hat{C}_3$ symmetric: $D_n(\hat{C}_3\mathbf r)=e^{i\frac{2\pi}{3}M(n)}D_n(\mathbf r)$, where the $\hat{C}_3$ quantum number $M(n)=\tau M(l)$ has opposite value in two valleys because of the time reversal relation. $M(l)=\{0,\pm1\}$ as a function of the band index $l$ is summarized in Table \ref{bands_C3_quantum_number}. Then the hopping integral satisfies
\begin{align}
&T_{nn'}(\hat{C}_3\mathbf R'-\hat{C}_3\mathbf R) \nonumber\\
&=\int D^*_n(\mathbf r-\hat{C}_3\mathbf R)\hat H_tD_{n'}(\mathbf r-\hat{C}_3\mathbf R')d\mathbf r \nonumber\\
&=\int D^*_n(\hat{C}_3\mathbf r-\hat{C}_3\mathbf R)\hat H_tD_{n'}(\hat{C}_3\mathbf r-\hat{C}_3\mathbf R')d\mathbf r \nonumber\\
&=e^{i\frac{2\pi}{3}(M(n')-M(n))}T_{nn'}(\mathbf R'-\mathbf R).\nonumber
\end{align}
With the equation above, applying Fourier transformation to $T_{nn'}(\mathbf r)$ results in
\begin{align}
t_{nn'}(\hat{C}_3\mathbf q)=&\frac{1}{\sqrt{\Omega\Omega'}}\int T_{nn'}(\mathbf r)e^{i\hat{C}_3\mathbf q\cdot\mathbf r}d\mathbf r \nonumber\\
=&\frac{1}{\sqrt{\Omega\Omega'}}\int T_{nn'}(\hat{C}_3\mathbf r)e^{i\hat{C}_3\mathbf q\cdot\hat{C}_3\mathbf r}d\mathbf r \nonumber\\
=&e^{i\frac{2\pi}{3}(M(n')-M(n))}t_{nn'}(\mathbf q).
\label{c3sym}
\end{align}
In the last step in Eq.~(\ref{c3sym}), we have used the relation $\hat{C}_3\mathbf q\cdot\hat{C}_3\mathbf r=\mathbf q\cdot\mathbf r$.

\subsection{In-plane mirror symmetry}
We use $\hat\sigma_\mathrm{v}$ to denote the mirror reflection operation on a real space vector $\mathbf{r}=(r_x,r_y)$ over the vertical $yz$ plane, i.e., $\hat\sigma_\mathrm{v}\mathbf{r}=(-r_x,r_y)$, or on a wave vector $\mathbf q=(q_x,q_y)$ as $\hat\sigma_\mathrm{v}\mathbf q=(-q_x,q_y)$. Obviously $\hat\sigma_\mathrm{v}\mathbf K=-\mathbf K$, thus under the mirror reflection $\psi_{\tau,0,n}(\hat\sigma_\mathrm{v}\mathbf r)=\psi_{-\tau,0,n}(\mathbf r)=\psi^*_{\tau,0,n}(\mathbf r)$, where the last step comes from the time reversal relation between the two valleys. Together with Eq. (\ref{Bloch1}), the local wavefunction $D_n(\mathbf r-\mathbf R)$ satisfies the property $D_n(\hat\sigma_\mathrm{v}\mathbf r-\hat\sigma_\mathrm{v}\mathbf R)=D_n^*(\mathbf r-\mathbf R)$. When both the upper and lower layer have the $yz$-plane mirror symmetry, i.e., R-stacking ($\theta=0^\circ$) or H-stacking ($\theta=60^\circ$), one gets
\begin{align}
&T_{nn'}(\hat\sigma_\mathrm{v}\mathbf R'-\hat\sigma_\mathrm{v}\mathbf R) \nonumber\\
&=\int D^*_n(\mathbf r-\hat\sigma_\mathrm{v}\mathbf R)\hat H_tD_{n'}(\mathbf r-\hat\sigma_\mathrm{v}\mathbf R')d\mathbf r \nonumber\\
&=\int D^*_n(\hat\sigma_\mathrm{v}\mathbf r-\hat\sigma_\mathrm{v}\mathbf R)\hat H_tD_{n'}(\hat\sigma_\mathrm{v}\mathbf r-\hat\sigma_\mathrm{v}\mathbf R')d\mathbf r \nonumber\\
&=T^*_{nn'}(\mathbf R'-\mathbf R).\nonumber
\end{align}

A Fourier transformation of $T_{nn'}(\mathbf r)$ results in
\begin{align}
t_{nn'}(\hat\sigma_\mathrm{v}\mathbf q)=&\frac{1}{\sqrt{\Omega\Omega'}}\int T_{nn'}(\mathbf r)e^{i\hat\sigma_\mathrm{v}\mathbf q\cdot\mathbf r}d\mathbf r \nonumber\\
=&\frac{1}{\sqrt{\Omega\Omega'}}\int T_{nn'}(\hat\sigma_\mathrm{v}\mathbf r)e^{i\hat\sigma_\mathrm{v}\mathbf q\cdot\hat\sigma_\mathrm{v}\mathbf r}d\mathbf r \nonumber\\
=&t^*_{nn'}(-\mathbf q).
\label{msym}
\end{align}
In the last step in Eq.~(\ref{msym}), we have used the relation $\hat\sigma_\mathrm{v}\mathbf q\cdot\hat\sigma_\mathrm{v}\mathbf r=\mathbf q\cdot\mathbf r$. Therefore, $t_{nn'}(\mathbf q)$ is real when $q_y=0$ in an R-type or H-type bilayer.

\section{$\pm\mathbf K$-valley coupling strength in H- and R-type homobilayers}

In homobilayer TMDs, the conduction and valence bands of the structures will be two-fold degenerate at $\tau\mathbf K$ point (without considering the spin-orbit coupling) if there is no interlayer coupling, i.e., $E_{c+j}=E_{c'+j}$, $E_{v-j}=E_{v'-j}$ with $j=0,1,2,\cdots$. The presence of the interlayer coupling will cause a finite energy level splitting $\Delta E_{c(v)}$, which contains the information of the hopping terms $t_{nn'}(\mathbf q)$.

We consider R-type ($\theta=0^\circ$) or H-type ($\theta=60^\circ$) TMD homobilayer structures with varying $\mathbf r_0$. As the two layers are fully commensurate, the interlayer hopping between $\tau\mathbf K$ in the lower layer and $\tau'\mathbf K'$ in the upper layer is allowed when $\tau'=\tau$ for R-stacking, and $\tau'=-\tau$ for H-stacking. To simplify the notation, we write $|n'\rangle\equiv|n',0\rangle$ and $|n\rangle\equiv|n,0\rangle$. Using Eq. (\ref{c3sym}) and (\ref{msym}), the hopping integral of Eq. (\ref{Hopping3}) between $\tau\mathbf K$ and $\tau'\mathbf K'$ can be written as
\begin{align}
&e^{-i\tau\mathbf K\cdot\mathbf r_0}\langle n|\hat H_t|n'\rangle
=\sum_{\bm\upkappa}t_{nn'}(\tau\bm\upkappa)e^{-i\tau\bm\upkappa\cdot\mathbf r_0} \nonumber\\
\approx&~\left(e^{-i\tau\mathbf K\cdot\mathbf r_0}+e^{-i\tau\hat{C}_3\mathbf K\cdot\mathbf r_0}e^{i\frac{2\pi}{3}(M(n')-M(n))}\right. \nonumber\\
&~\left.+e^{-i\tau\hat{C}^2_3\mathbf K\cdot\mathbf r_0}e^{i\frac{4\pi}{3}(M(n')-M(n))}\right)t^{(0)}_{nn'} \nonumber\\
+&~\left(e^{2i\tau\mathbf K\cdot\mathbf r_0}+e^{2i\tau\hat{C}_3\mathbf K\cdot\mathbf r_0}e^{i\frac{2\pi}{3}(M(n')-M(n))}\right. \nonumber\\
&~\left.+e^{2i\tau\hat{C}^2_3\mathbf K\cdot\mathbf r_0}e^{i\frac{4\pi}{3}(M(n')-M(n))}\right)t^{(1)}_{nn'} \nonumber\\
+&~\left(e^{-i\tau\bm\upkappa_1\cdot\mathbf r_0}+e^{-i\tau\hat{C}_3\bm\upkappa_1\cdot\mathbf r_0}e^{i\frac{2\pi}{3}(M(n')-M(n))}\right. \nonumber\\
&~\left.+e^{-i\tau\hat{C}^2_3\bm\upkappa_1\cdot\mathbf r_0}e^{i\frac{4\pi}{3}(M(n')-M(n))}\right)t^{(2)}_{nn'} \nonumber\\
+&~\left(e^{-i\tau\bm\upkappa_2\cdot\mathbf r_0}+e^{-i\tau\hat{C}_3\bm\upkappa_2\cdot\mathbf r_0}e^{i\frac{2\pi}{3}(M(n')-M(n))}\right. \nonumber\\
&~\left.+e^{-i\tau\hat{C}^2_3\bm\upkappa_2\cdot\mathbf r_0}e^{i\frac{4\pi}{3}(M(n')-M(n))}\right)(t^{(2)}_{nn'})^*.
\label{hopping_commensurate_bilayer}
\end{align}
Here, $t^{(0)}_{nn'}\equiv t_{nn'}(\tau\mathbf K)$ corresponds to the main hopping term, $t^{(1)}_{nn'}\equiv t_{nn'}(-2\tau\mathbf K)$ is the $1$st order term, and $t^{(2)}_{nn'}\equiv t_{nn'}(\tau\bm\upkappa_1)=t^*_{nn'}(\tau\bm\upkappa_2)$ is the $2$nd order term. Note that $t^{(0)}_{nn'}$ and $t^{(1)}_{nn'}$ are real due to Eq. (\ref{msym}), while $t^{(2)}_{nn'}$ is complex. We have dropped the other higher order terms with larger $|\bm\upkappa|$ whose contributions are expected to be negligible.

Now we analyze the conduction band splitting $\Delta E_c$. Because of the large splitting between two different bands, the hopping between lower layer $c$-band and upper layer $n'$-band with $n'\ne c'$ can be well accounted by a second-order perturbation, which results in an energy shift $\delta E_c(\mathbf r_0)\equiv\sum_{n'\ne c'}\frac{|\langle c|\hat H_t|n'\rangle|^2}{E_c-E_{n'}}$ to the $c$-band. Similarly, the lower layer $n$-band with $n\ne c$ results in an energy shift $\delta E_{c'}(\mathbf r_0)\equiv\sum_{n\ne c}\frac{|\langle c'|\hat H_t|n\rangle|^2}{E_{c'}-E_n}$ to the $c'$-band. So in the subspace spanned by $|c'\rangle$ and $|c\rangle$, the hopping Hamiltonian has a form
\begin{align}
\hat H_{cc'}=&\left(E_c+\delta E_c(\mathbf r_0)\right)|c\rangle\langle c|+\left(E_c+\delta E_{c'}(\mathbf r_0)\right)|c'\rangle\langle c'| \nonumber\\
&+\langle c'|\hat H_t|c\rangle|c'\rangle\langle c|+h.c..
\end{align}
$\Delta E_c$ is then given by the energy splitting between the eigenstates of $\hat H_{cc'}$, which is
\begin{align}
\Delta E_c=\sqrt{\left(\delta E_c(\mathbf r_0)-\delta E_{c'}(\mathbf r_0)\right)^2+4|\langle c'|\hat H_t|c\rangle|^2}.
\end{align}
The same analysis can be applied to the valence bands, which gives
\begin{align}
\Delta E_v=\sqrt{\left(\delta E_v(\mathbf r_0)-\delta E_{v'}(\mathbf r_0)\right)^2+4|\langle v'|\hat H_t|v\rangle|^2},
\end{align}
with $\delta E_v(\mathbf r_0)\equiv\sum_{n'\ne v'}\frac{|\langle v|\hat H_T|n'\rangle|^2}{E_v-E_{n'}}$ and $\delta E_{v'}(\mathbf r_0)\equiv\sum_{n\ne v}\frac{|\langle v'|\hat H_T|n\rangle|^2}{E_{v'}-E_n}$.

For most of the $\mathbf r_0$ values, the corresponding R- or H-type commensurate bilayer structures are unstable thus don't exist in nature. However, these structures can locally exist in an incommensurate bilayer with large scale moir\'{e} superlattice pattern.\cite{Tong2016arXiv,MacDonald2016arXiv,STMS} In a local region with a size much larger than the monolayer lattice constant but much smaller than the moir\'{e} supercell, the atomic registry between the two layers is locally indistinguishable from an R- or H-type commensurate bilayer, which is characterized by a continuously varying $\mathbf r_0$. The local band structure of this region is then given by that of the commensurate bilayer with the corresponding $\mathbf r_0$ value.\cite{Tong2016arXiv,MacDonald2016arXiv,STMS}
As $\mathbf r_0$ varies from position to position in a moir\'{e} supercell, the $\mathbf r_0$-dependent conduction/valence band energy shifts $\delta E_{c/v}(\mathbf r_0)$ can be responsible for the observed position-dependent local band gap modulation. \cite{STMS}

\subsection{H-type homobilayer}
\begin{figure}[]
\includegraphics[width=\linewidth]{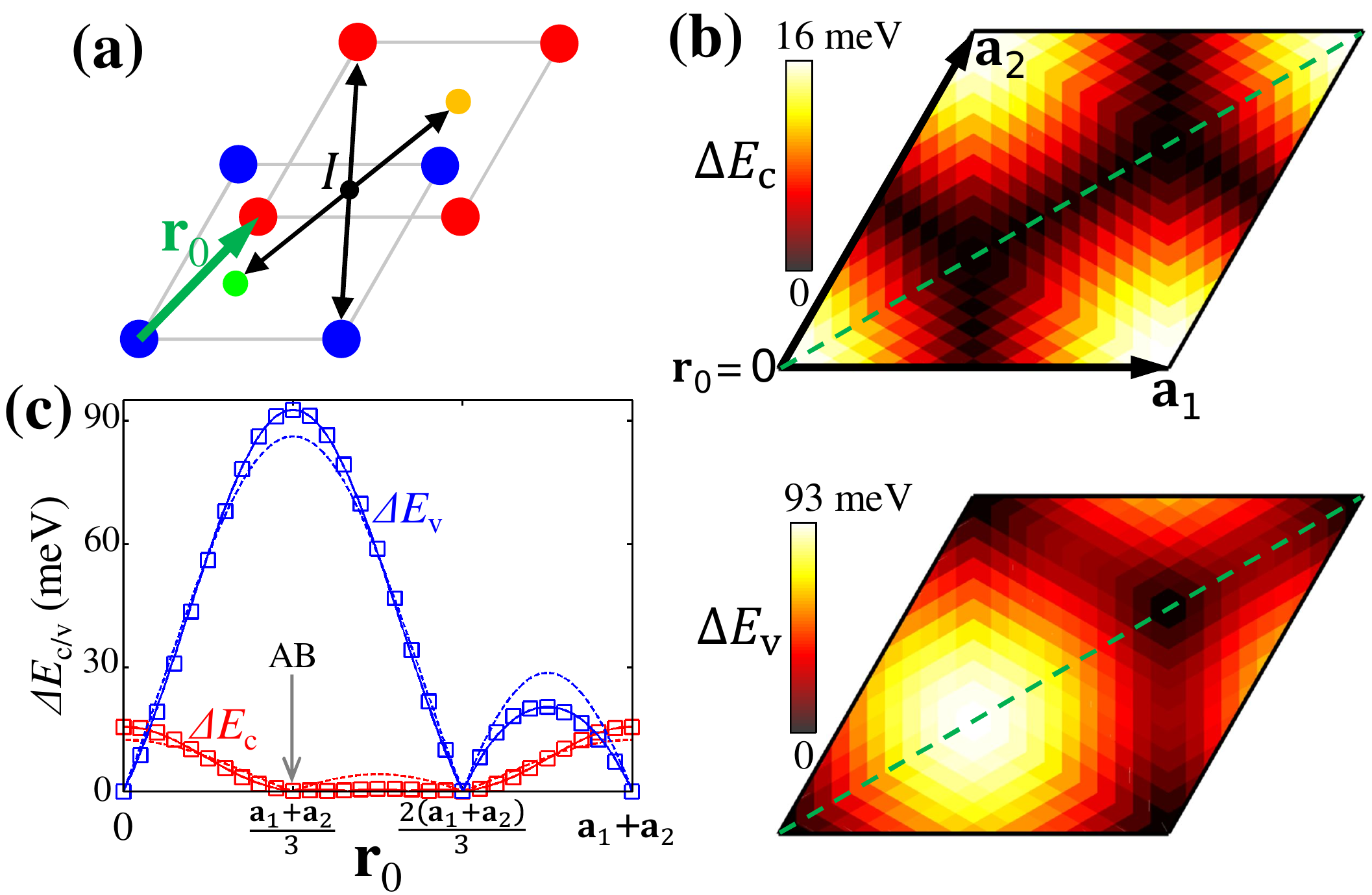}
\caption{(Color online) (a) An H-type TMD homobilayer with any interlayer translation $\mathbf r_0$ always has an inversion center $I$. Here, the large red (blue) dots denote the metal atoms in the upper (lower) layer, and the small orange (green) dots denote the chalcogen atoms in the upper (lower) layer. (b) The \textit{ab initio} results for $\Delta E_{c/v}$ as functions of $\mathbf r_0$. (c) 2D plots showing the $\Delta E_{c/v}$ line cuts along dashed green lines in (b), where the symbols are the \textit{ab initio} results and the solid curves are the fits using $t^{(0)}_{cc'/vv'}$, $t^{(1)}_{cc'/vv'}$ and $t^{(2)}_{cc'/vv'}$. The dashed curves are the results keeping only the main terms $t^{(0)}_{cc'/vv'}$. The natural TMD homobilayers with AB stacking correspond to $\mathbf r_0=(\mathbf a_1+\mathbf a_2)/3$.}
\label{Fig2}
\end{figure}

For H-type stacking, the two states with finite hopping strength in different layers have the opposite valley indices $\tau=-\tau'$. Using Eq. (\ref{hopping_commensurate_bilayer}) together with the $M(n)$ values given in Table \ref{bands_C3_quantum_number}, we find $\delta E_c(\mathbf r_0)=\delta E_{c'}(\mathbf r_0)$ and $\delta E_v(\mathbf r_0)=\delta E_{v'}(\mathbf r_0)$. This can be understood from the symmetry consideration. As shown in Fig. \ref{Fig2}(a), an H-type homobilayer with an arbitrary $\mathbf r_0$ has a spatial inversion center, which means the two layers are symmetric. So $\delta E_{c/v}(\mathbf r_0)$, the lower layer conduction/valence band energy shift induced by the remote bands in the upper layer, is always equivalent to $\delta E_{c'/v'}(\mathbf r_0)$ which is the upper layer energy shift induced by the lower layer. The band splittings are then simply given by
\begin{align}
\Delta E_c=2|\langle c'|\hat H_t|c\rangle|, ~~~~\Delta E_v=2|\langle v'|\hat H_t|v\rangle|.
\label{H_type_splitting}
\end{align}

We have performed \textit{ab initio} calculations for the band structures of MoS$_2$ H-type homobilayers with different $\mathbf r_0$. For each given $\mathbf r_0$, we fix the interlayer distance defined as the vertical distance between the nearest chalcogen atoms of neighboring layers at $d=2.975$ {\AA} (the experimental bulk value \cite{bulk_d}), and the other lattice parameters are taken from Ref. [\onlinecite{LatticeParameter}]. The energy splitting values $\Delta E_{c/v}$ are calculated with the projector-agumented wave (PAW) method implemented in the Quantum Espresso package.\cite{QEspresso} The Perdew-Burke-Ernzerhof (PBE) exchange-correlation functional and scalar relativistic pseudopotential without including the spin-orbit coupling has been exploited, and the cutoff energy for plane wave basis is set as $80$ Ry. A $15\times15\times1$ $\mathbf k$-point sample is generated by the Monkhorst-Pack (MP) approach, and the self-consistent ground state is achieved with the total energy converge criteria $10^{-10}$ Ry.

The calculation results are presented in Fig. \ref{Fig2}(b) as surface plots. In the 2D plot of Fig. \ref{Fig2}(c) with $\mathbf r_0$ along the long diagonal line of the unit cell, we show both the \textit{ab initio} results (symbols) and the corresponding fits (solid lines) using Eq. (\ref{H_type_splitting}) and (\ref{hopping_commensurate_bilayer}) by keeping the $t^{(0)}_{cc'/vv'}$, $t^{(1)}_{cc'/vv'}$ and $t^{(2)}_{cc'/vv'}$ hopping terms. The two show excellent agreement. The dashed lines are the results keeping only the main hopping terms $t^{(0)}_{cc'/vv'}$, which can already reproduce the major features. Thus those $t_{nn'}(\tau\bm\upkappa)$ with larger $|\tau\bm\upkappa|$ are indeed negligible. The fitting parameters are summarized in Table \ref{H_type_hop_str}, which give $|t^{(0)}_{cc'}|\gg|t^{(1)}_{cc'}|, |t^{(2)}_{cc'}|$ and $|t^{(0)}_{vv'}|\gg|t^{(1)}_{vv'}|, |t^{(2)}_{vv'}|$ as we expected. So in H-type commensurate bilayers, it is a good approximation to write the $\mathbf K$-point conduction/valence band interlayer couplings in the forms
\begin{align}
\label{hopping_H_type}
&|\langle c|\hat H_t|c'\rangle_H|
\approx\left|e^{i\mathbf K\cdot\mathbf r_0}+e^{i\hat{C}_3\mathbf K\cdot\mathbf r_0}+e^{i\hat{C}^2_3\mathbf K\cdot\mathbf r_0}\right|t^{(0)}_{cc'},\\
&|\langle v|\hat H_t|v'\rangle_H|
\approx\left|e^{i\mathbf K\cdot\mathbf r_0}+e^{i(\hat{C}_3\mathbf K\cdot\mathbf r_0+\frac{2\pi}{3})}+e^{i(\hat{C}^2_3\mathbf K\cdot\mathbf r_0+\frac{4\pi}{3})}\right|t^{(0)}_{vv'}. \nonumber
\end{align}
The above equations should also apply to H-type commensurate heterobilayers. Similar forms have been obtained in early papers \cite{Tong2016arXiv,Bistritzer2011PNAS,SanJose2014PRB,Jung2014PRB}. Here we would like to point out that, in these interlayer coupling forms the $e^{\pm i2\pi/3}$ phase factors have different origins for bilayer TMD and graphene systems. In bilayer TMDs it is from the $\hat C_3$ quantum number $M(n)$ of the atomic orbital combination $D_n$, as clearly indicated by Eq. (\ref{c3sym}) and (\ref{hopping_commensurate_bilayer}). While in bilayer Graphene, it originates from the displacement vectors between the nearest A and B sublattice sites.

From the above equations, we get $t^{(0)}_{cc'}\approx\Delta E_c/6$ at $\mathbf r_0=0$ and $t^{(0)}_{vv'}\approx\Delta E_v/6$ at $\mathbf r_0=(\mathbf a_1+\mathbf a_2)/3$ for H-type homobilayers.

\begin{table}[htbp]
\begin{center}
\caption{The obtained hopping strengths for the H-type homobilayer MoS$_2$ from fitting to the \textit{ab initio} results of band splitting. The main hopping term $t^{(0)}_{vv'}$ is consistent with our previous result.\cite{Tong2016arXiv}}
\label{H_type_hop_str}
\begin{tabular}{cccccc}
\hline\hline
$t^{(0)}_{cc'}$ & $t^{(1)}_{cc'}$ & $|t^{(2)}_{cc'}|$ & $t^{(0)}_{vv'}$ & $t^{(1)}_{vv'}$ & $|t^{(2)}_{vv'}|$ \\
\hline
$2.1$~meV & $0.4$~meV & $0.1$~meV & $14.4$~meV & $1.2$~meV & $0.4$~meV \\
\hline\hline
\end{tabular}
\end{center}
\end{table}

\subsection{R-type homobilayer}
\begin{figure}[]
\includegraphics[width=\linewidth]{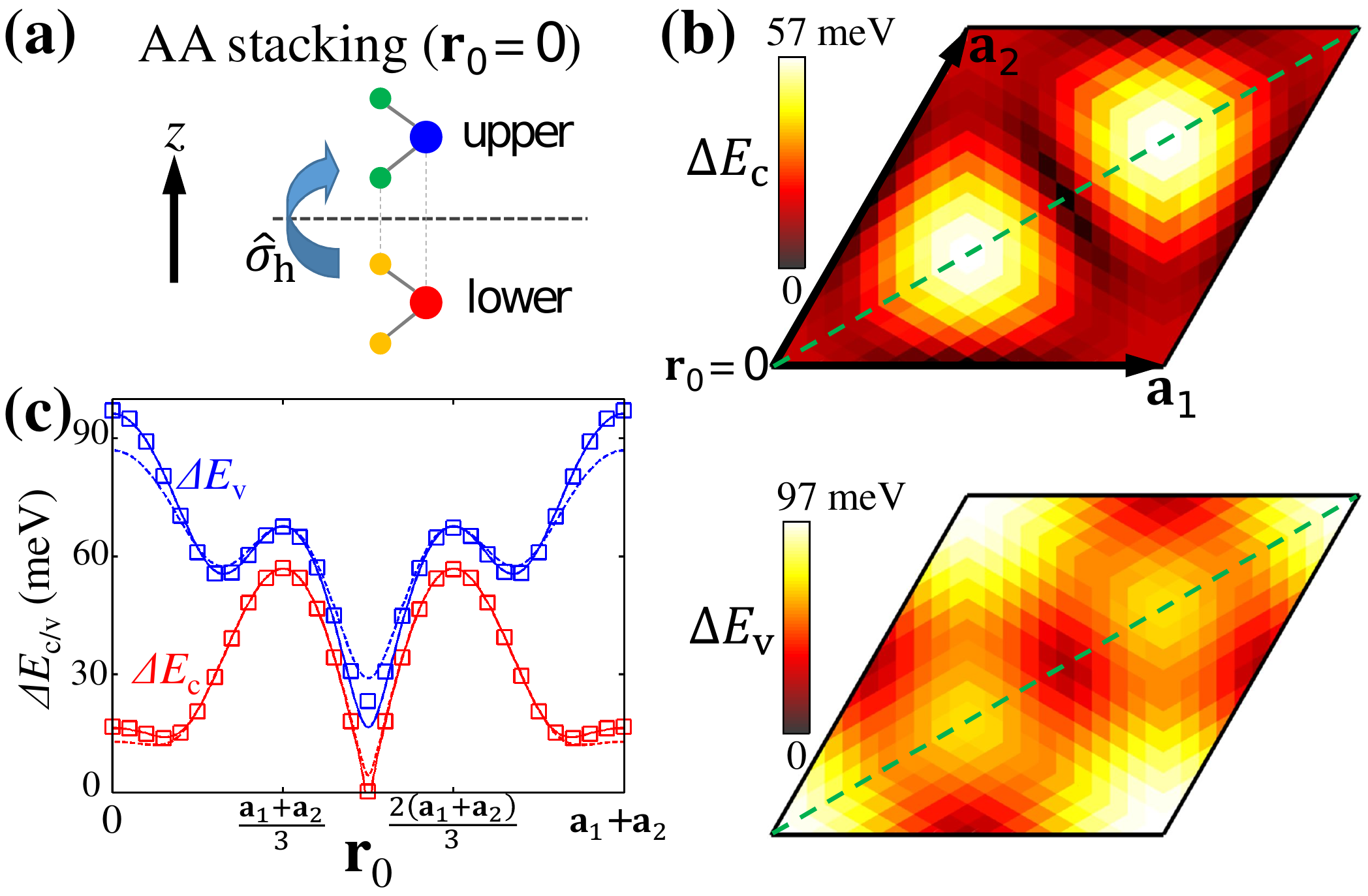}
\caption{(Color online) (a) An AA-type TMD homobilayer has an out-of-plane mirror reflection ($\hat\sigma_\textrm{h}$) symmetry. (b) The \textit{ab initio} results for $\Delta E_{c/v}$ as functions of $\mathbf r_0$ for R-type homobilayer MoS$_2$. (c) 2D plots showing the $\Delta E_{c/v}$ line cuts along dashed green lines in (b), where the symbols are the \textit{ab initio} results and the solid curves are the fits using $\delta E^{(0)}_{c/v}$, $t^{(0)}_{cc'/vv'}$ and $t^{(1)}_{cc'/vv'}$. The dashed curves are the results keeping only the main hopping terms $\delta E^{(0)}_{c/v}$ and $t^{(0)}_{cc'/vv'}$.}
\label{Fig3}
\end{figure}

In contrast to the H-stacking, the R-type homobilayer is not inversion symmetric thus generally the upper and lower layers are not equivalent. We find $\delta E_c(\mathbf r_0)\ne\delta E_{c'}(\mathbf r_0)$ and $\delta E_v(\mathbf r_0)\ne\delta E_{v'}(\mathbf r_0)$ for a general R-type stacking. Only for AA staking with $\mathbf r_0=0$, which has the out-of-plane mirror reflection ($\hat\sigma_\textrm{h}$) symmetry (Fig. \ref{Fig3}(a)), the two layers become equivalent and $\delta E_{c/v}(\mathbf r_0=0)=\delta E_{c'/v'}(\mathbf r_0=0)$.

For R-type stacking, the two states with finite hopping strength in different layers have the same valley indices $\tau=\tau'$. Using Eq. (\ref{hopping_commensurate_bilayer}) together with the $M(n)$ values given in Table \ref{bands_C3_quantum_number}, we can write the $\mathbf r_0$-dependence of $\delta E_{c/v}(\mathbf r_0)$ and $\delta E_{c'/v'}(\mathbf r_0)$ as
\begin{align}
\delta E_c(\mathbf r_0)-\delta E_{c'}(\mathbf r_0)\approx&~\delta E^{(0)}_{c}f(\mathbf r_0), \nonumber\\
\delta E_v(\mathbf r_0)-\delta E_{v'}(\mathbf r_0)\approx&~\delta E^{(0)}_{v}f(\mathbf r_0),
\end{align}
where $\delta E^{(0)}_{c/v}$ are from the main hopping terms: 
\begin{align}
\label{band_shift}
\delta E^{(0)}_c\equiv&\frac{|t^{(0)}_{c,v'}|^2}{E_c-E_v}-\frac{|t^{(0)}_{c,c'+1}|^2}{E_{c+1}-E_c}-\frac{|t^{(0)}_{c,v'-1}|^2}{E_c-E_{v-1}} \\
&-\frac{|t^{(0)}_{c,v'-3}|^2}{E_c-E_{v-3}}+\frac{|t^{(0)}_{c,c'+2}|^2}{E_{c+2}-E_c}+\frac{|t^{(0)}_{c,c'+3}|^2}{E_{c+3}-E_c}\cdots, \nonumber\\
\delta E^{(0)}_v\equiv&\frac{|t^{(0)}_{v,v'-1}|^2}{E_v-E_{v-1}}+\frac{|t^{(0)}_{v,v'-3}|^2}{E_v-E_{v-3}}-\frac{|t^{(0)}_{v,c'+2}|^2}{E_{c+2}-E_v} \nonumber\\
&-\frac{|t^{(0)}_{v,c'+3}|^2}{E_{c+3}-E_v}+\frac{|t^{(0)}_{v,c'}|^2}{E_c-E_{v}}-\frac{|t^{(0)}_{v,v'-2}|^2}{E_v-E_{v-2}}\cdots, \nonumber
\end{align}
and other higher order terms are ignored. So for R-type homobilayer TMDs, the conduction/valence band splitting has a form
\begin{align}
\Delta E_c=\sqrt{\left(\delta E^{(0)}_cf(\mathbf r_0)\right)^2+4|\langle c'|\hat H_t|c\rangle|^2}, \nonumber\\
\Delta E_v=\sqrt{\left(\delta E^{(0)}_vf(\mathbf r_0)\right)^2+4|\langle v'|\hat H_t|v\rangle|^2},
\label{R_type_splitting}
\end{align}
where
\begin{align}
f(\mathbf r_0)&\equiv\left|e^{i\mathbf K\cdot\mathbf r_0}+e^{i(\hat{C}_3\mathbf K\cdot\mathbf r_0+\frac{2\pi}{3})}+e^{i(\hat{C}^2_3\mathbf K\cdot\mathbf r_0+\frac{4\pi}{3})}\right|^2 \nonumber\\
&-\left|e^{i\mathbf K\cdot\mathbf r_0}+e^{i(\hat{C}_3\mathbf K\cdot\mathbf r_0-\frac{2\pi}{3})}+e^{i(\hat{C}^2_3\mathbf K\cdot\mathbf r_0-\frac{4\pi}{3})}\right|^2. \nonumber
\end{align}

For $\mathbf r_0=0$, $f(\mathbf r_0)=0$ and $\Delta E_c=2|\langle c'|\hat H_t|c\rangle|$, $\Delta E_v=2|\langle v'|\hat H_t|v\rangle|$, which agrees with our symmetry analysis that the two layers of AA stacking are related by $\hat\sigma_\textrm{h}$ and thus are equivalent. On the other hand, $\langle c'|\hat H_t|c\rangle=\langle v'|\hat H_t|v\rangle=0$ for $\mathbf r_0=\pm\frac{\mathbf a_1+\mathbf a_2}{3}$, which leads to $\Delta E_c=9|\delta E^{(0)}_{c}|$ and $\Delta E_v=9|\delta E^{(0)}_{v}|$.

We have also performed \textit{ab initio} calculations for $\Delta E_{c/v}$ in MoS$_2$ R-type homobilayers with different $\mathbf r_0$. The calculation details are the same as in the H-type case, and the results are presented in Fig. \ref{Fig3}(b). Once gain we show both the \textit{ab initio} results (symbols) and the corresponding fits (solid lines) using Eq. (\ref{R_type_splitting}) and (\ref{hopping_commensurate_bilayer}) in Fig. \ref{Fig3}(c) with $\mathbf r_0$ along the long diagonal line of the unit cell. Keeping only the main hopping terms $\delta E^{(0)}_{c/v}$ and $t^{(0)}_{cc'/vv'}$ (dashed lines) can already reproduce the major features, whereas the fits using $\delta E^{(0)}_{c/v}$, $t^{(0)}_{cc'/vv'}$ and $t^{(1)}_{cc'/vv'}$ terms (solid lines) agree almost perfectly with the \textit{ab initio} results. The fitting parameters are summarized in Table \ref{R_type_hop_str}. As a good approximation, the $\mathbf K$ point conduction/valence band interlayer hoppings in R-type commensurate bilayers take the forms
\begin{align}
&|\langle c|\hat H_t|c'\rangle_R|
\approx\left|e^{i\mathbf K\cdot\mathbf r_0}+e^{i\hat{C}_3\mathbf K\cdot\mathbf r_0}+e^{i\hat{C}^2_3\mathbf K\cdot\mathbf r_0}\right|t^{(0)}_{cc'}, \nonumber\\
&|\langle v|\hat H_t|v'\rangle_R|
\approx\left|e^{i\mathbf K\cdot\mathbf r_0}+e^{i\hat{C}_3\mathbf K\cdot\mathbf r_0}+e^{i\hat{C}^2_3\mathbf K\cdot\mathbf r_0}\right|t^{(0)}_{vv'}.
\label{hopping_R_type}
\end{align}
The equations above also apply to R-type commensurate heterobilayers. Similar forms have been obtained in early papers.\cite{Tong2016arXiv,Bistritzer2011PNAS,SanJose2014PRB,Jung2014PRB}

From the above equations, we get $t^{(0)}_{cc'}\approx\Delta E_c/6$ and $t^{(0)}_{vv'}\approx\Delta E_v/6$ at $\mathbf r_0=0$ for R-stacking. 

\begin{table}[htbp]
\begin{center}
\caption{The obtained hopping strengths for the R-type homobilayer MoS$_2$ from fitting to the \textit{ab initio} results of band splitting. The main hopping term $t^{(0)}_{vv'}$ is consistent with our previous result.\cite{Tong2016arXiv}}
\label{R_type_hop_str}
\begin{tabular}{cccccc}
\hline\hline
$|\delta E^{(0)}_c|$ & $t^{(0)}_{cc'}$ & $t^{(1)}_{cc'}$ & $|\delta E^{(0)}_v|$ & $t^{(0)}_{vv'}$ & $t^{(1)}_{vv'}$ \\
\hline
$6.3$~meV & $2.1$~meV & $0.6$~meV & $7.5$~meV & $14.5$~meV & $1.6$~meV \\
\hline\hline
\end{tabular}
\end{center}
\end{table}

\subsection{Variation of coupling strength with interlayer distance}

As shown in both the theoretical analysis above and the good fit results in Fig. \ref{Fig2}(c) and \ref{Fig3}(c), $t^{(j)}_{nn'}$ doesn't directly depend on the interlayer translation $\mathbf r_0$. However, $t^{(j)}_{nn'}$ should sensitively depend on the interlayer distance $d$, the equilibrium value of which varies in a large range depending on the stacking pattern in R- or H-type commensurate bilayers characterized by $\mathbf r_0$. \cite{Liu2014NC,vanderZande2014NL} A recent scanning tunneling microscopy/spectroscopy experiment has shown that in a single heterobilayer structure with the formation of large scale moir\'{e} superlattice, $d$ can vary from position to position due to the variation of local stacking patterns.\cite{STMS}

We have calculated these $\Delta E_{c/v}$ as functions of $d$, which can be well fitted by exponential functions $\Delta E_n(d)=\Delta E_{0,n}e^{-d/d_n}$. Here, $\Delta E_{0,c}=1.96$~eV, $d_{c}=0.62$~{\AA}, and $\Delta E_{0,v}=14.4$~eV, $d_{v}=0.59$~{\AA} for R-stacking, $\Delta E_{0,c}=1.77$~eV, $d_{c}=0.63$~{\AA}, and $\Delta E_{0,v}=12.7$~eV, $d_{v}=0.61$~{\AA} for H-stacking.

\begin{figure}[]
\includegraphics[width=\linewidth]{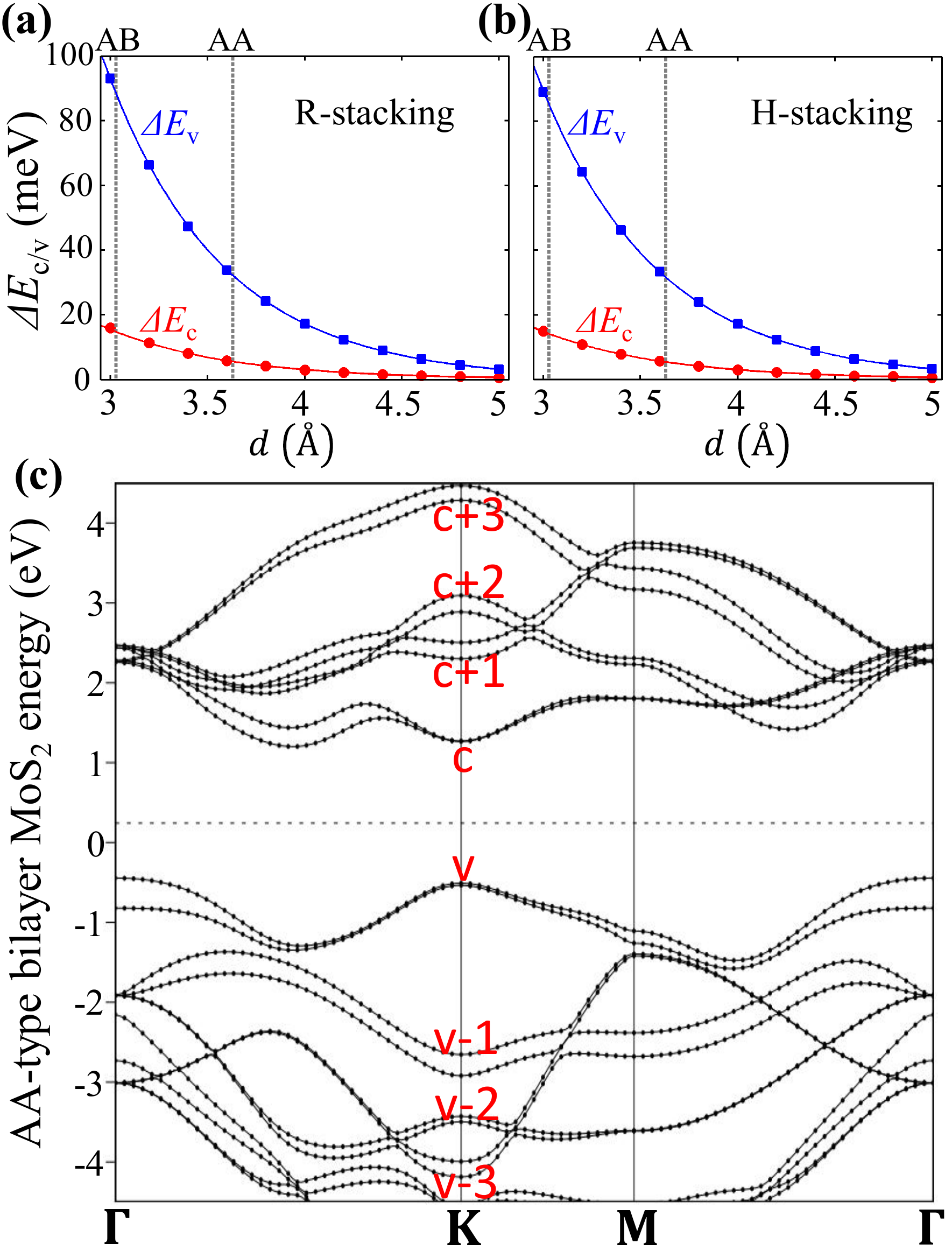}
\caption{(Color online) (a) The solid dots are our \textit{ab initio} results of conduction/valence band splitting at $\mathbf K$-point for MoS$_2$ R-type homobilayers with $\mathbf r_0=0$ (AA stacking) as functions of interlayer distance $d$ (defined as the vertical distance between the nearest chalcogen atoms of neighboring layers). The solid curves are the exponential fits. These results are also presented in Ref. [\onlinecite{Tong2016arXiv}]. The vertical dashed lines show the numerical values of interlayer distance for AB ($3.0$ {\AA}) and AA ($3.6$ {\AA}) homobilayer MoS$_2$, adopted from Ref. [\onlinecite{Liu2014NC}]. (b) The case for MoS$_2$ H-type homobilayers. $\Delta E_c$ is for $\mathbf r_0=0$, while $\Delta E_v$ is for $\mathbf r_0=(\mathbf a_1+\mathbf a_2)/3$ (AB stacking). (c) The \textit{ab initio} band structure of an AA-type MoS$_2$ homobilayer without spin-orbit coupling, where the band splittings $\Delta E_n$ can be clearly seen. The interlayer distance is set as $d_\textrm{AA}=3.72$ {\AA}, and the other calculation details are the same as those in Fig. \ref{Fig2} and \ref{Fig3}.}
\label{Fig4}
\end{figure}

Considering the similarity of the d-orbitals of Mo and W atoms, the hopping strengths for the homobilayers shall provide reasonable estimations to those in the TMD heterobilayers. However, in heterobilayers the $\pm\mathbf K$ valleys have much larger conduction/valence band offsets, which leads to negligible layer mixing.\cite{MoS2WSe2_offset,MoSe2WSe2_offset} For example, in MoS$_2$/WSe$_2$ heterobilayer, the $\pm\mathbf K$-valley valence (conduction) band offset is found to be $0.83$~eV ($0.76$~eV).\cite{MoS2WSe2_offset} While in MoSe$_2$/WSe$_2$ heterobilayer, the valence band offset is $0.3$~eV. \cite{MoSe2WSe2_offset} These values are all much larger than the $\pm\mathbf K$ valley coupling strengths which are on the order of several tens of meV. Thus unlike the homobilayers where $t_{cc'/vv'}$ should be treated nonperturbatively, in heterobilayers all $t_{cn'/vn'}$ hopping terms can be treated perturbatively.

\subsection{Interlayer coupling strengths of other bands}

Just like the $c$ and $v$ bands, the band splitting values $\Delta E_{c+j}$ ($\Delta E_{v-j}$) of other bands in an AA-type homobilayer give the corresponding interlayer coupling strengths $t^{(0)}_{c+j,c'+j}$ ($t^{(0)}_{v-j,v'-j}$). We show the band structure of an AA-type homobilayer MoS$_2$ in Fig. \ref{Fig4}(c). The extracted $\Delta E_n$ for $n$ from $c+3$ to $v-3$ are summarized in Table \ref{AA_band_split}. Note that some bands have much larger interlayer coupling than those of the $c$ and $v$ bands, which could be related to their larger $p$-orbital proportions in the corresponding orbital combinations $D_n$ (see Table \ref{bands_C3_quantum_number}). It is then natural to expect that the hopping terms $t^{(0)}_{cn'}$ ($t^{(0)}_{vn'}$) between the remote band $n'$ and the conduction band $c$ (valence band $v$) are much larger than $t^{(0)}_{cc'}$ ($t^{(0)}_{vv'}$), which can lead to a large range modulation of the $c$ and $v$ band energy with $\mathbf r_0$. This is consistent with the observed $\sim 0.1$ eV band gap modulation in a TMD heterobilayer \cite{STMS}.

\begin{table}[htbp]
\begin{center}
\caption{The band splittings $\Delta E_n$ (in unit of meV) for $n$ from $c+3$ to $v-3$ extracted from Fig. \ref{Fig4}(c).}
\label{AA_band_split}
\begin{tabular}{cccccccc}
\hline\hline
 $c+3$ & $c+2$ & $c+1$ & $c$ & $v$ & $v-1$ & $v-2$ & $v-3$ \\
\hline
 ~~$183$~~ & ~~$206$~~ & ~~$203$~~ & ~~$5$~~ & ~~$27$~ & ~~$265$~~ & ~~$67$~~ & ~~$193$~~ \\
\hline\hline
\end{tabular}
\end{center}
\end{table}

\section{Twisted or lattice-mismatched bilayer structures and moir\'{e} patterns}

The interlayer couplings in twisted or lattice-mismatched bilayers can also be described by Eq. (\ref{Hopping3}). Note that the local atomic orbital $D_{n'}(\mathbf r)$ in the upper layer is rotated by the interlayer twist angle $\theta$ with respect to $D_n(\mathbf r)$ in the lower layer. So in principle the corresponding hopping term $t_{nn'}$ for $\theta\ne0^\circ$ and $60^\circ$ should be different from those given in the previous discussion of H-type or R-type commensurate bilayers. However, when considering the cases with close to $0^\circ$ or $60^\circ$ twist angle, it is a good approximation to replace $t_{nn'}$ by those of the H-type or R-type commensurate bilayers.

\begin{figure}[]
\includegraphics[width=\linewidth]{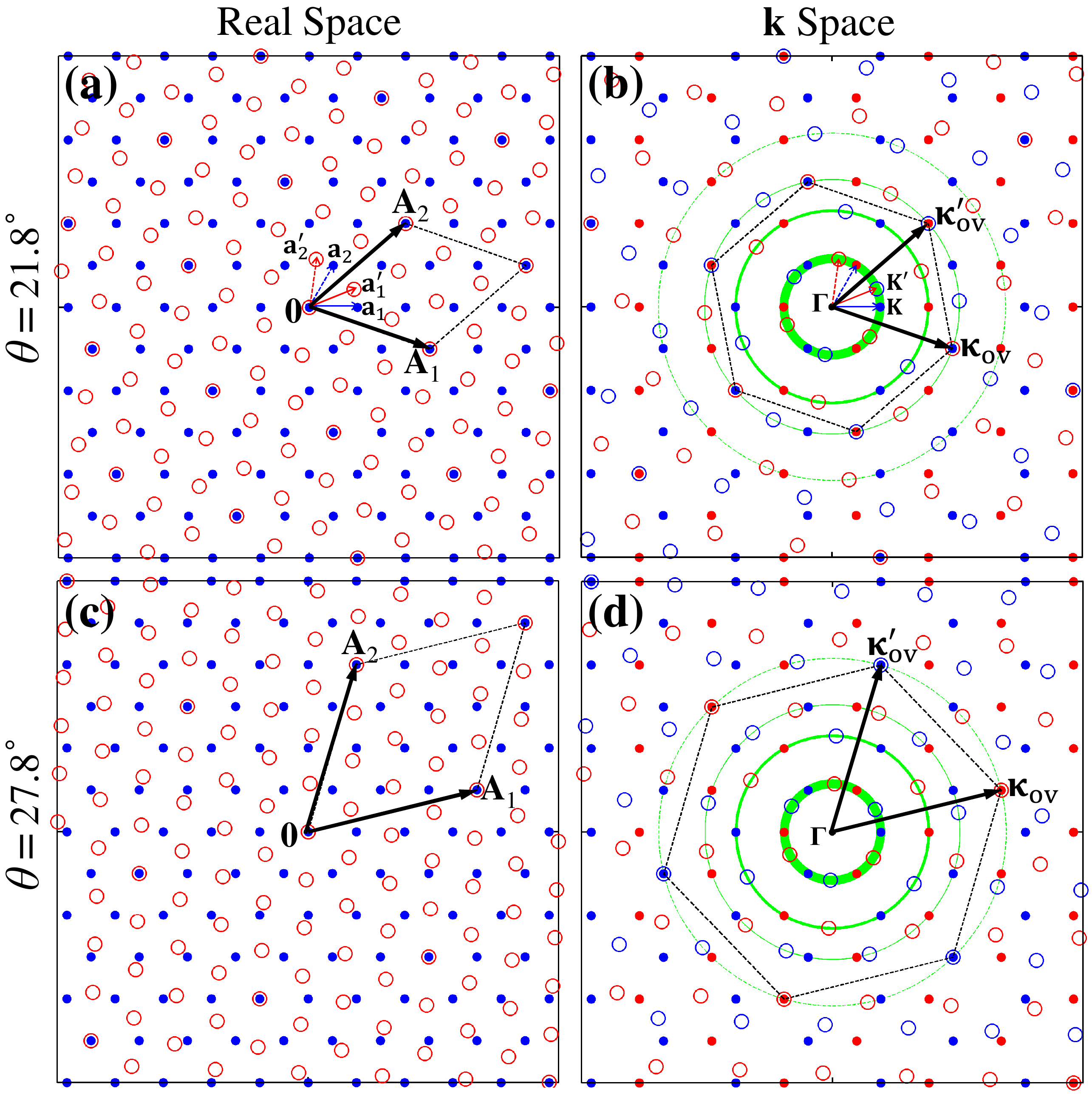}
\caption{(Color online) (a) The real space atomic registry of a lattice-matched commensurate bilayer with $\theta=21.8^\circ$ twist angle. The solid blue (empty red) dots denote the metal atoms in the lower (upper) layer, with $\mathbf a_{1,2}$ ($\mathbf a'_{1,2}$) the corresponding unit lattice vectors. The rhombus corresponds to a supercell, with its four vertices located at $0$, $\mathbf A_1$, $\mathbf A_2$ and $\mathbf A_1+\mathbf A_2$ where two metal atoms in opposite layers horizontally overlap. (b) The corresponding $\mathbf k$-space configurations of the two layers. The solid blue (red) dots correspond to $\bm\upkappa$ ($-\bm\upkappa$) in the lower layer, and the empty blue (red) dots correspond to $\bm\upkappa'$ ($-\bm\upkappa'$) in the upper layer. The six overlapping $(\tau\bm\upkappa,\tau'\bm\upkappa')$ pairs on the third smallest green circle form a hexagon (dashed lines). The $\mathbf k$-space hexagon corner $\bm\upkappa_\textrm{ov}$ ($\bm\upkappa'_\textrm{ov}$) corresponds to $\mathbf A_1$ ($\mathbf A_2$) in the real space. (c) and (d) Another commensurate bilayer with twist angle $\theta=27.8^\circ$ with larger supercell size and $|\bm\upkappa_\textrm{ov}|$.}
\label{Fig5}
\end{figure}

A twisted or lattice-mismatched bilayer can still be commensurate under special conditions, i.e., the two layers form a periodic superlattice structure with the supercell size larger than the monolayer unit cell. As the commensurability is irrelevant to the interlayer translation, we assume a metal atom in the upper layer horizontally overlaps with a metal atom in the lower layer at the $xy$-plane origin. In the commensurate case, the bilayer supercell is then given by the smallest rhombus with its four vertices located at the overlapping metal atoms, as shown in Fig. \ref{Fig5}(a). Notice that in $\mathbf k$-space, $\tau\bm\upkappa$ in the lower layer overlaps with $\tau'\bm\upkappa'$ in the upper layer at certain positions $\bm\upkappa_\textrm{ov}$, which means $\tau\mathbf K$ and $\tau'\mathbf K'$ are coupled through the interlayer hopping $t_{nn'}(\bm\upkappa_\textrm{ov})$ (see Eq. (\ref{Hopping3})). Interestingly, there is one-to-one correspondence between the superlattice unit vector $\mathbf A_{1,2}$ and $\bm\upkappa_\textrm{ov}$ such that $|\bm\upkappa_\textrm{ov}|=\frac{4\pi}{3aa'}|\mathbf A_{1,2}|$, as shown in Fig. \ref{Fig5}(b). So larger supercell size corresponds to larger $|\bm\upkappa_\textrm{ov}|$ and thus smaller coupling strength $|t_{nn'}(\bm\upkappa_\textrm{ov})|$ between $\tau\mathbf K$ and $\tau'\mathbf K'$, which agrees with the findings in a recent work.\cite{Zhou2016arXiv} In fact, in a twisted or lattice-mismatched commensurate bilayer, $\bm\upkappa_\textrm{ov}$ always corresponds to second or higher order hopping terms $t^{(j)}_{nn'}$ ($j\ge2$), which are negligibly small compared to the main terms $t^{(0)}_{nn'}$.

Away from the band edges $\tau\mathbf K$ and $\tau'\mathbf K'$, the interlayer coupling can be significant. We can always find small wave vectors $\mathbf k'$ and $\mathbf k$ where $\mathbf k'-\mathbf k$ equals to $\tau\mathbf K-\tau'\mathbf K'$ or $\tau\hat C_3\mathbf K-\tau'\hat C_3\mathbf K'$ or $\tau\hat C^2_3\mathbf K-\tau'\hat C^2_3\mathbf K'$. According to Eq. (\ref{Hopping3}), the coupling between $\tau\mathbf K+\mathbf k$ and $\tau'\mathbf K'+\mathbf k'$ is then $\sim t^{(0)}_{nn'}$, which corresponds to the main hopping term. Note that such coupling terms are insensitive to whether the bilayer is commensurate or not. As discussed above, in a twisted or lattice-mismatched bilayer the commensurability only introduces direct coupling between the two band edges $\tau\mathbf K$ and $\tau'\mathbf K'$, with a negligibly small coupling strength.

\begin{figure}[]
\includegraphics[width=\linewidth]{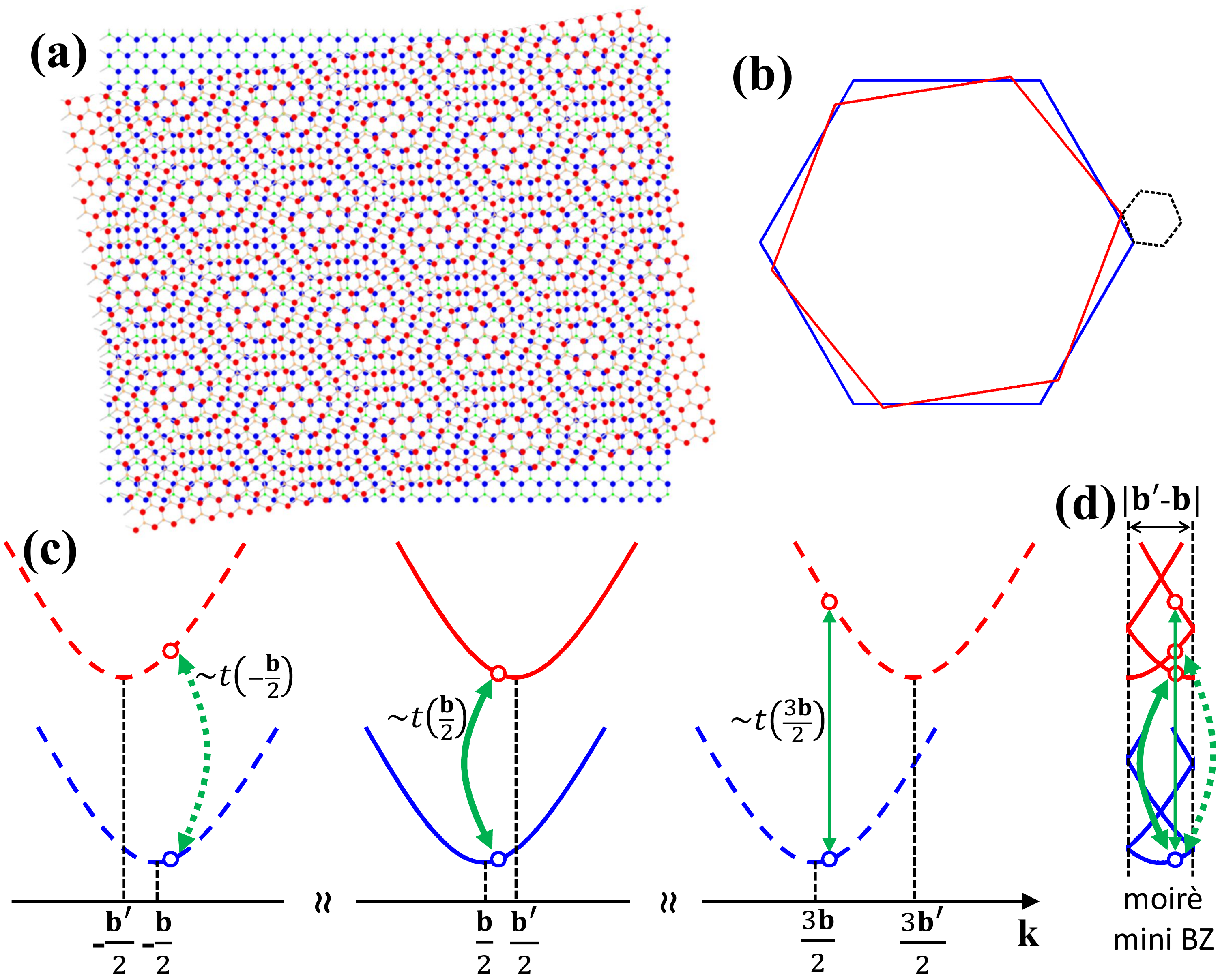}
\caption{(Color online) (a) A typical hexagonal bilayer moir\'{e} pattern with lattice constant mismatch $\delta=0.05$ and twist angle $\theta=9^\circ$. (b) The monolayer BZs (solid blue and red hexagons) and the moir\'{e} superlattice mini BZ (dashed black hexagon). (c) The band dispersions of two 1D systems (denoted as blue and red colors). Only those near the band edges located at $\mathbf b/2$ and $\mathbf b'/2$ are shown. The double arrows indicate the $t(j\mathbf b+\mathbf b/2+\mathbf k)$ hopping terms with $j=0,\pm1$ (Eq. (\ref{1Dhopping})), and the arrow thickness corresponds to the hopping strength. (d) The corresponding interlayer hopping terms (double arrows) between different mini bands in the 1D moir\'{e} mini BZ.}
\label{Fig6}
\end{figure}

The interlayer coupling between $\tau\mathbf K+\mathbf k$ and $\tau'\mathbf K'+\mathbf k'$ discussed above is especially important for bilayers with $\tau\mathbf K$ and $\tau'\mathbf K'$ close to each other, in which $|\mathbf k|$ and $|\mathbf k'|$ can be small enough that low energy carriers in different layers are efficiently coupled. On the other hand, it is known that in such a bilayer with $|\tau\mathbf K-\tau'\mathbf K'|\ll4\pi/3a$, a moir\'{e} superlattice pattern with large scale periodicity will form,\cite{Ponomarenko2013Nature,Dean2013Nature,Hunt2013Science,Jung2015NC,Jung2014PRB} as shown in Fig. \ref{Fig6}(a). Below we show that the moir\'{e} superlattice picture is fully consistent with our theoretical analysis in Section II.

We note that the moir\'{e} pattern is not a rigorous periodic structure but a good approximation, whose emergence can be understood as follows. Any quantity involving the periodicity of both layers (e.g., $\psi^*_{n,\mathbf k}\psi_{n',\mathbf k'}$ which appears in the hopping integral in Eq. (\ref{Hopping1})) can be written as the sum of all $e^{i(\mathbf G-\mathbf G')\cdot\mathbf r}$ terms by a Fourier transformation. Here, $\mathbf G=j_1\mathbf b_1+j_2\mathbf b_2$ ($\mathbf G'=j'_1\mathbf b'_1+j'_2\mathbf b'_2$) are the lower (upper) layer reciprocal lattice vectors, with $\mathbf b_{1,2}$ ($\mathbf b'_{1,2}$) the corresponding primitive reciprocal lattice vectors and $j_{1,2}$, $j'_{1,2}$ integers. Those terms with large $|\mathbf G|$ or $|\mathbf G'|$ are related to the fast oscillating components in $\psi_{n,\mathbf k}$ or $\psi_{n',\mathbf k'}$ with periods much smaller than the lattice constant, and can be dropped. Then the remaining slowly oscillating terms always have $\mathbf G-\mathbf G'=j_1(\mathbf b_1-\mathbf b'_1)+j_2(\mathbf b_2-\mathbf b'_2)$. Thus the large scale moir\'{e} period is characterized by the primitive reciprocal lattice vectors $\mathbf B_1\equiv\mathbf b_1-\mathbf b'_1$ and $\mathbf B_2\equiv\mathbf b_2-\mathbf b'_2$. The above analysis requires $|\mathbf B_{1,2}|\approx\frac{4\pi}{\sqrt{3}a}\sqrt{\delta^2+\delta\theta^2}\ll\frac{4\pi}{\sqrt{3}a}$, with $\delta=a/a'-1$ and $\delta\theta$ the twist angle deviation to $0$ or $\pi/3$. The moir\'{e} superlattice constant is then $A\approx a/\sqrt{\delta^2+\delta\theta^2}\gg a$, with $|\delta|\ll1$ and $|\delta\theta|\ll1$ the prerequisites for the existence of a moir\'{e} pattern.

The moir\'{e} superlattice mini Brillouin zone (BZ) has its corners located at $\tau\mathbf K-\tau'\mathbf K'$ and its $\pi/3$ rotations (see Fig. \ref{Fig6}(b)).\cite{Bistritzer2011PNAS,Jung2014PRB,SanJose_PRB2014} The mini BZ forms a complete basis in $\mathbf k$-space for the bilayer structure. Inside the mini BZ, the original monolayer bands are folded into a series of closely spaced mini bands, and a state with wave vector $\mathbf k$ in one layer can hop to various mini bands in the other layer with the same $\mathbf k$. We note that for small $|\bm\upkappa|$ and $|\bm\upkappa'|$, the delta function in Eq. (\ref{Hopping3}) can be written as $\delta_{\mathbf k-\mathbf k',\tau\bm\upkappa-\tau'\bm\upkappa'}=\delta_{(\tau\mathbf K+\mathbf k)-(\tau'\mathbf K'+\mathbf k'),j_1\mathbf B_1+j_2\mathbf B_2}$, which is just the momentum conservation condition in the mini BZ picture. The effect of the interlayer coupling is to open gaps between the mini bands.

For convenience, we use two 1D systems to illustrate the correspondence between the individual BZs and the moir\'{e} mini BZ in Fig. \ref{Fig6}(c) and \ref{Fig6}(d). The band edges of the individual 1D systems are assumed to be located at $\mathbf b/2$ and $\mathbf b'/2$, where $\mathbf b$ and $\mathbf b'$ are the primitive reciprocal lattice vectors of the corresponding systems and $|\mathbf b'-\mathbf b|\ll |\mathbf b|$. Following Eq. (\ref{Bloch2}) the Bloch states can be written as
\begin{align}
\psi_{\mathbf b/2+\mathbf k}(\mathbf r)=&\frac{1}{\sqrt{N}}\sum_{\mathbf R}e^{i(\mathbf b/2+\mathbf k)\cdot\mathbf R}D(\mathbf r-\mathbf R), \nonumber\\
\psi_{\mathbf b'/2+\mathbf k'}(\mathbf r)=&\frac{1}{\sqrt{N'}}\sum_{\mathbf R'}e^{i(\mathbf b'/2+\mathbf k')\cdot\mathbf R'}D(\mathbf r-\mathbf R').
\end{align}
Here $D(\mathbf r-\mathbf R)$ is the atomic orbital combination localized near $\mathbf R$. Analogous to Eq. (\ref{Hopping3}), we write the hopping integral between the two 1D systems as
\begin{align}
&\int\psi^*_{\mathbf b/2+\mathbf k}(\mathbf r)\hat H_t\psi_{\mathbf b'/2+\mathbf k'}(\mathbf r)d\mathbf r \nonumber\\
&=\sum_{jj'}\delta_{(j+\frac{1}{2})\mathbf b+\mathbf k,(j'+\frac{1}{2})\mathbf b'+\mathbf k'}t(j\mathbf b+\mathbf b/2+k)e^{-ij\mathbf b\cdot\mathbf r_0}.
\label{1Dhopping}
\end{align}
Here, $t(j\mathbf b+\mathbf b/2+\mathbf k)$ with $j=0,\pm1,\cdots$ are the Fourier transformations of the hopping integral between the two localized orbitals, which are indicated as double arrows near $(j+1/2)\mathbf b$ in Fig. \ref{Fig6}(c). These terms with different $j$ have one-to-one correspondence with those between different mini bands in the moir\'{e} mini BZ, as shown in Fig. \ref{Fig6}(d). 

Although the individual BZ picture is equivalent to the moir\'{e} mini BZ as discussed above, we find that it is more convenient to extract the hopping strength using the former picture. Considering that the magnitude of $t(\mathbf q)$ decays fast with the increase of $|\mathbf q|$, in the individual BZ picture we can just focus on the hopping terms $t(\mathbf q)$ with $\mathbf q$ inside the first BZs. Whereas in the moir\'{e} mini BZ picture we cannot directly get which two mini bands have a strong hopping strength. 

On the other hand, a local picture becomes more convenient for describing large scale moir\'{e} superlattices.\cite{Tong2016arXiv,MacDonald2016arXiv,STMS} We can consider a local region with a size much larger than the monolayer unit cell, but at the same time much smaller than the moir\'{e} supercell. The corresponding atomic registry is locally indistinguishable from an R- or H-type commensurate bilayer, thus we can discuss its local band structure which is given by that of the corresponding commensurate bilayer. Different local regions are characterized by different $\mathbf r_0$, which results in a periodic modulation of the local band structure. In TMD heterobilayers where the $\pm\mathbf K$ valleys have negligible layer mixing, the interlayer coupling appears as a local band structure modulation, which is equivalent to applying band-dependent external superlattice potentials on two decoupled layers.\cite{MacDonald2016arXiv}

\section{Interlayer coupling in $\bm\Gamma_\textrm{v}$ and $\mathbf Q_\textrm{c}$ valleys}

\begin{figure}[]
\includegraphics[width=\linewidth]{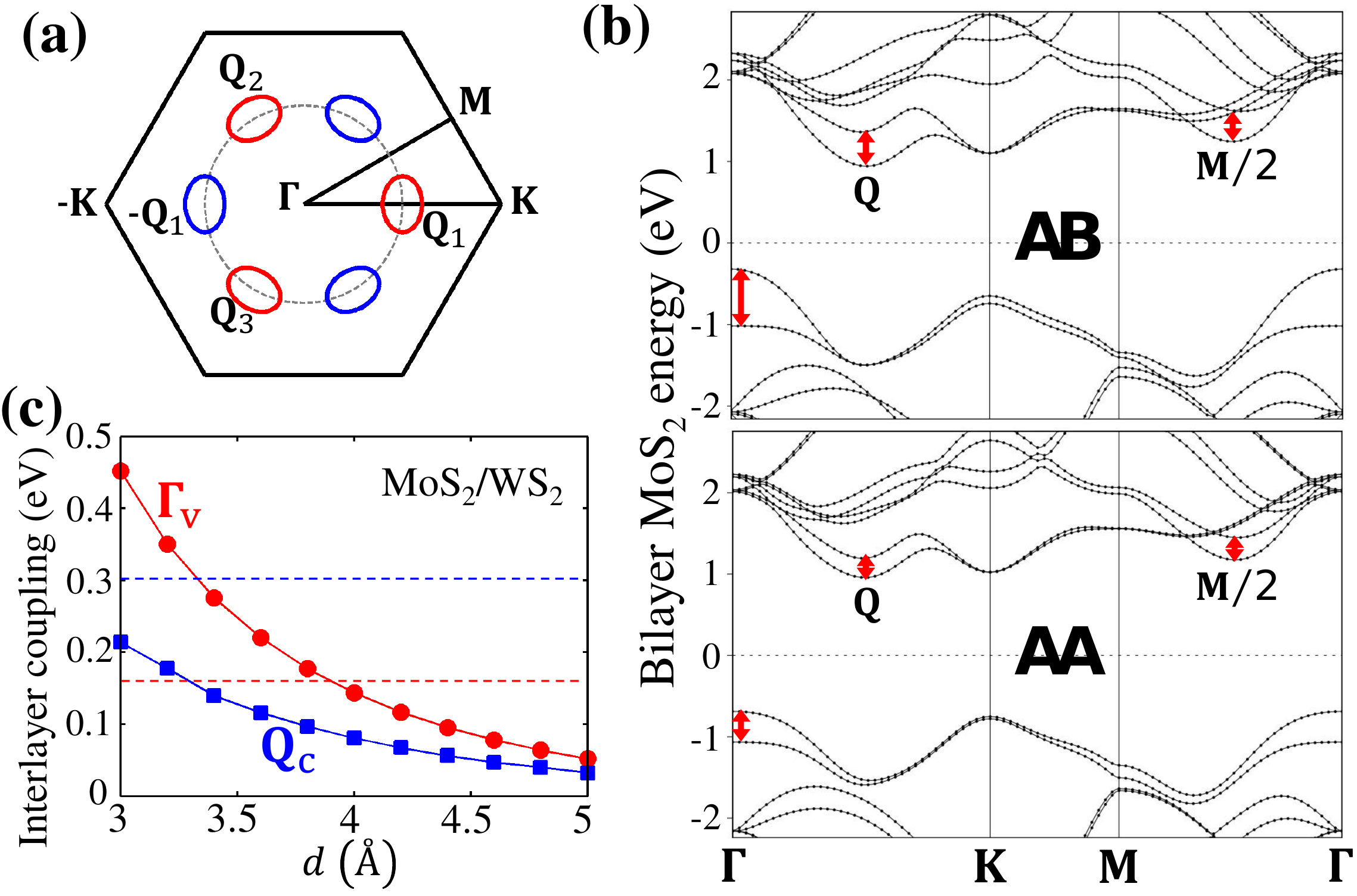}
\caption{(Color online) (a) The red and blue pockets illustrate the energy contours of the six $\mathbf Q_\textrm{c}$ valleys. The dashed circle corresponds to a ring-shaped region with strong conduction band interlayer coupling. (b) The \textit{ab initio} band structures of AB and AA homobilayer MoS$_2$ with interlayer distances $d_\textrm{AB}=2.975$ {\AA} and $d_\textrm{AA}=3.72$ {\AA}, respectively. The spin-orbit coupling is not considered. The band splittings at $\mathbf \Gamma_\textrm{v}$, $\mathbf Q_\textrm{c}$, and $\mathbf M/2$ are denoted by red arrows. (c) The obtained interlayer coupling strength at $\bm\Gamma_\textrm{v}$ and $\mathbf Q_\textrm{c}$ for AA-type MoS$_2$/WS$_2$ heterobilayers. The red (blue) dashed line shows the interlayer band offset $\Delta E_{0,\bm\Gamma}=0.16$ eV ($\Delta E_{0,\mathbf Q}=0.3$ eV).}
\label{Fig7}
\end{figure}

In 2H homobilayer TMDs, the $\bm\Gamma_\textrm{v}$ and $\mathbf Q_\textrm{c}$ energies are strongly shifted away from the corresponding monolayer positions as evidenced by the photoluminescence and ARPES measurements,\cite{Liu2014NC,vanderZande2014NL,Yeh2016NL} which is a signature of the strong interlayer coupling near these positions.\cite{Liu2015CSR} Here, $\bm\Gamma_\textrm{v}$ denotes the $\bm\Gamma$ point of valence band, and $\mathbf Q_\textrm{c}$ denotes the six conduction band extrema near the middle of the $\bm\Gamma$-$\tau\mathbf K$ lines (Fig. \ref{Fig7}(a)). From the \textit{ab initio} results of homobilayer band structures, we estimate that the interlayer hopping strengths in the $\bm\Gamma_\textrm{v}$ and $\mathbf Q_\textrm{c}$ valleys are on the order of several hundred meV (Fig. \ref{Fig7}(b)).

Note that all $\mathbf Q_\textrm{c}$ points are located on a ring with radius $\sim|\mathbf K|/2$ (Fig. \ref{Fig7}(a)), while Fig. \ref{Fig7}(b) indicates a strong interlayer coupling near the conduction band $\mathbf M/2$ point (the middle of the $\bm\Gamma$-$\mathbf M$ line). Thus we speculate that all conduction band $\mathbf k$ points on this ring have strong interlayer couplings. Furthermore, for an arbitrary interlayer twist angle, the $\bm\Gamma$ positions are not affected and the $\mathbf Q_\textrm{c}$ valleys are always on this ring. Therefore, we expect that the interlayer twist does not change the strong coupling nature of $\bm\Gamma_\textrm{v}$ and $\mathbf Q_\textrm{c}$ valleys.

We have also calculated the band splitting $\Delta E_{\bm\Gamma}$ ($\Delta E_{\mathbf Q}$) at $\bm\Gamma_\textrm{v}$ ($\mathbf Q_\textrm{c}$) point for AA-type MoS$_2$/WS$_2$ heterobilayers, which is found to depend sensitively on the interlayer distance $d$. The band splitting can be approximated as $\Delta E_{\bm\Gamma/\mathbf Q}=\sqrt{(\Delta E_{0,\bm\Gamma/\mathbf Q})^2+4t^2_{\bm\Gamma/\mathbf Q}}$, when ignoring the coupling with
other bands. For a large enough $d$, i.e., under the vanishing interlayer coupling limit, the interlayer hopping strength $t_{\bm\Gamma/\mathbf Q}\to0$, from which we get the band offset values $\Delta E_{0,\mathbf Q}=0.3$ eV and $\Delta E_{0,\bm\Gamma}=0.16$ eV. $t_{\bm\Gamma/\mathbf Q}$ for intermediate values of $d$ are then derived from the relation above and shown in Fig. \ref{Fig7}(c).

The $\bm\Gamma_\textrm{v}$ and $\mathbf Q_\textrm{c}$ valley Bloch functions can be approximated similar to Eq.~(\ref{Bloch2}) for the $\pm\mathbf K$ valley
\begin{align}
\psi_{\bm\Gamma,\mathbf k}(\mathbf r)\approx&\frac{1}{\sqrt{N}}\sum_{\mathbf R}e^{i\mathbf k\cdot\mathbf R}D_{\bm\Gamma}(\mathbf r-\mathbf R), \nonumber\\
\psi_{\tau\mathbf Q_j,\mathbf k}(\mathbf r)\approx&\frac{1}{\sqrt{N}}\sum_{\mathbf R}e^{i(\tau\mathbf Q_j+\mathbf k)\cdot\mathbf R}D_{\tau\mathbf Q}(\mathbf r-\mathbf R).
\label{GammaQBlochFunction}
\end{align}
Here, we use $\tau\mathbf Q_j$ with $\tau=\pm$ and $j=1,2,3$ to distinguish the six degenerate but inequivalent $\mathbf Q_\textrm{c}$ (Fig. \ref{Fig7}(a)), which are related by $\hat C_3$ or time reversal operations. $D_{\bm\Gamma}(\mathbf r-\mathbf R)$ and $D_{\tau\mathbf Q}(\mathbf r-\mathbf R)$ are the linear combinations of atomic orbitals localized around $\mathbf R$ for the corresponding valleys. Following the derivation of Eq. (\ref{Hopping3}), the hopping strength can be written as
\begin{eqnarray}
\langle\bm\Gamma,\mathbf k|\hat H_t|\bm\Gamma',\mathbf k'\rangle &\approx& t_{\bm\Gamma}(\mathbf k)\delta_{\mathbf k',\mathbf k}, \nonumber\\
\langle\tau\mathbf Q_j,\mathbf k|\hat H_t|\tau'\mathbf Q'_{j'},\mathbf k'\rangle
&\approx& t_{\mathbf Q}(\tau\mathbf Q_j+\mathbf k)\delta_{\tau'\mathbf Q'_{j'}+\mathbf k',\tau\mathbf Q_j+\mathbf k}. \nonumber
\end{eqnarray}
In the last step above, we have used the fact that $\mathbf G+\mathbf k$ and $\tau\mathbf Q_j+\mathbf G+\mathbf k$ are well outside the monolayer first BZ when $\mathbf G\ne0$, and the corresponding $t_{\bm\Gamma}(\mathbf G+\mathbf k)e^{i\mathbf G\cdot\mathbf r_0}$ and $t_{\mathbf Q}(\tau\mathbf Q_j+\mathbf G+\mathbf k)e^{i\mathbf G\cdot\mathbf r_0}$ terms have much smaller magnitudes than those at $\mathbf G=0$ and can be ignored. Thus, unlike the $\pm\mathbf K$ valleys (Eq. (\ref{hopping_H_type}) and (\ref{hopping_R_type})) discussed previously, the $\bm\Gamma_\textrm{v}$ and $\mathbf Q_\textrm{c}$ valley interlayer couplings are nearly independent of the interlayer translation $\mathbf r_0$.

The interlayer coupling strengths of $\bm\Gamma_\textrm{v}$ and $\mathbf Q_\textrm{c}$ valleys are comparable to the corresponding band offsets in TMD heterobilayers (Fig. \ref{Fig7}(c)), which is distinct from the $\pm\mathbf K$ valleys. The strong interlayer couplings of $\bm\Gamma_\textrm{v}$ and $\mathbf Q_\textrm{c}$ valleys originate from: (1) the non-ignorable $p_z$ orbital of chalcogen atoms \cite{Liu2015CSR} in $D_{\bm\Gamma}$ and $D_{\tau\mathbf Q}$; (2) the fact that they correspond to the $t(\mathbf q)$ Fourier components with $|\mathbf q|<|\mathbf K|$. The resulting strong layer mixing can play an important role in the interlayer charge transfer processes of TMD heterobilayers with type-II band alignments. \cite{Rivera_InterlayerX0,Fang_PNAS,Chiu_ACSNano,Lee_NatNano,Furchi_NL,Cheng_NL,Ceballos_2014ACSNano,Hong_NatNano,Yu_2015NL,Rigosi_2015NL}

\begin{figure}[t]
\includegraphics[width=\linewidth]{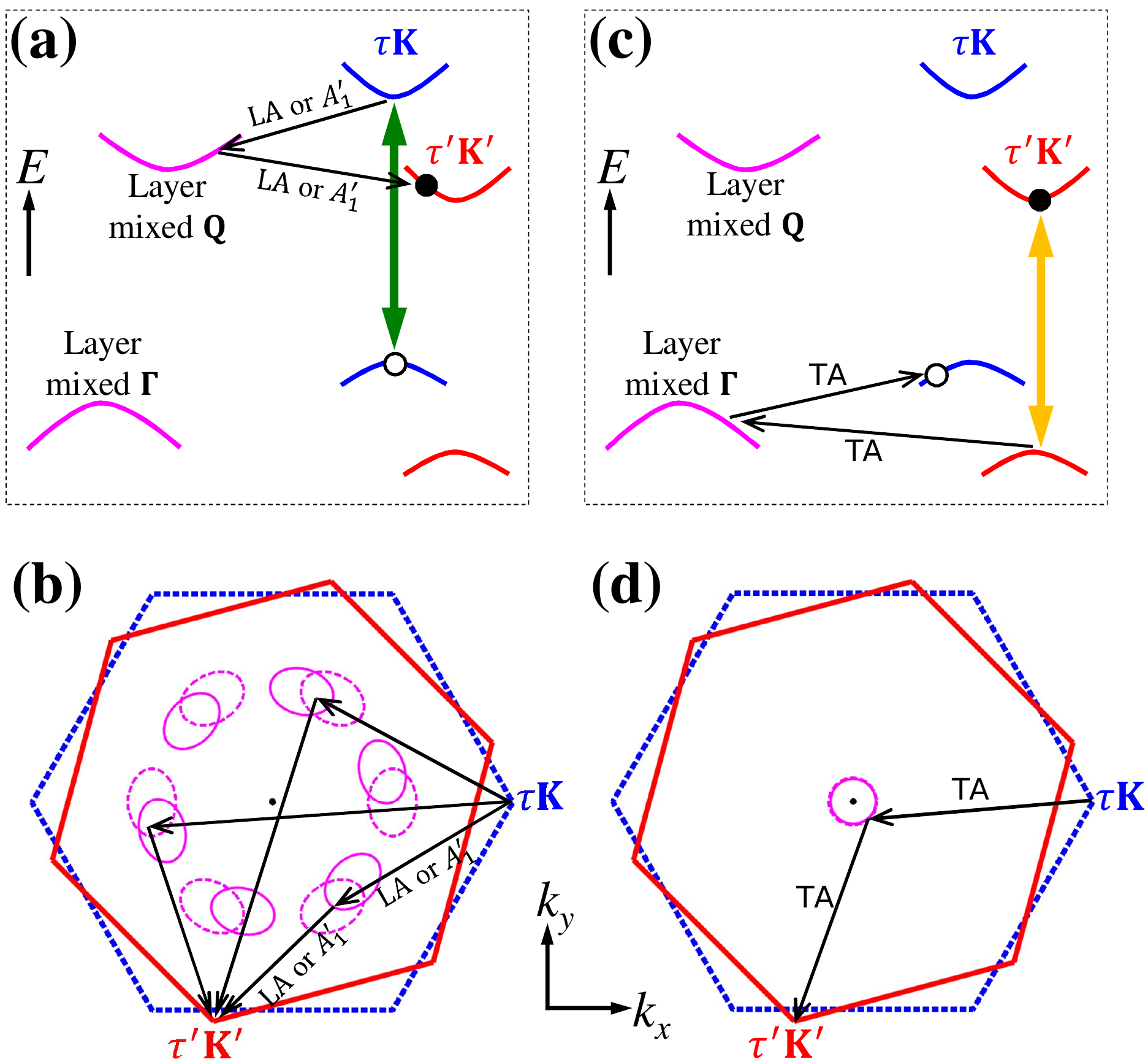}
\caption{(Color online) (a) Schematic illustration of the electron interlayer charge transfer process in the energy space. The blue (red) curves are the lower (upper) layer $\tau\mathbf K$ ($\tau'\mathbf K'$) valley bands, and the purple curves are the strongly layer mixed $\bm\Gamma_\textrm{v}$ and $\mathbf Q_\textrm{c}$ valleys. The double arrow illustrates the optical generation of electron-hole pairs in the $\tau\mathbf K$ valley. The single arrows correspond to the electron relaxation pathways. (b) The electron interlayer charge transfer process in the momentum space. The dashed blue (solid red) hexagon is the lower (upper) layer BZ. The electron can be scattered to three $\mathbf Q_\textrm{c}$ valleys through emitting a phonon with wave vector $\mathbf M$. (c)\&(d) Schematic illustration of the hole interlayer charge transfer.}
\label{Fig8}
\end{figure}

Experiments have found that the charge transfer process is ultrafast ($<50$ fs) and independent on the interlayer twist, \cite{Hong_NatNano,Yu_2015NL,Rigosi_2015NL} both of which cannot be explained by the weak interlayer coupling strength of the $\pm\mathbf K$ valleys. Here, we propose the following electron (hole) interlayer charge transfer mechanism mediated by the $\mathbf Q_\textrm{c}$ ($\bm\Gamma_\textrm{v}$) valley. For a type-II heterobilayer with the conduction (valence) band edge located at $\tau'\mathbf K'_\textrm{c}$ ($\tau\mathbf K_\textrm{v}$), a high energy electron in $\tau\mathbf K_\textrm{c}$ valley can relax to one of the $\mathbf Q_\textrm{c}$ valleys through scattering with phonons, other carriers or impurities/defects. As $\mathbf Q_\textrm{c}$ valleys are strongly layer mixed, this electron can further relax to the $\tau'\mathbf K'_\textrm{c}$ valley, as shown in Fig. \ref{Fig8}(a) and \ref{Fig8}(b). A high energy hole in $\tau'\mathbf K'_\textrm{v}$ can relax to the strongly layer mixed $\bm\Gamma_\textrm{v}$ valley and then to $\tau\mathbf K_\textrm{v}$, see Fig. \ref{Fig8}(c) and \ref{Fig8}(d). We expect that such interlayer charge transfer rate is close to the $\pm\mathbf K$ valley carrier relaxation rate in few-layer or bulk TMDs, since they both involve intervalley relaxation from $\pm\mathbf K$ to $\mathbf Q_\textrm{c}$ or $\bm\Gamma_\textrm{v}$. Actually the measured intervalley relaxation time in few-layer MoS$_2$ is $\sim20$ fs,\cite{Nie_2014ACSNano} which indeed agrees well with the interlayer charge transfer time ($<50$ fs) in heterobilayer TMDs. \cite{Hong_NatNano,Yu_2015NL,Rigosi_2015NL}

DFT calculations suggest that electrons in $\mathbf K_\textrm{c}$ valley couple strongly with LA and $A'_1$ phonons with wave vectors in the vicinity of $\mathbf M$,\cite{electron_phonon_coupling} which leads to scatterings between $\mathbf K_\textrm{c}$ and $-\mathbf Q_{1,2,3}$ valleys. On the other hand, holes in $\mathbf K_\textrm{v}$ valley couple strongly with TA phonons with wave vectors in the vicinity of $-\mathbf K$, \cite{hole_phonon_coupling} which leads to scatterings between $\mathbf K_\textrm{v}$ and $\bm\Gamma_\textrm{v}$ valleys. Using the Fermi golden rule, we can estimate the phonon emission assisted electron/hole intervalley scatterings rates as
\begin{align}
\frac{1}{\tau_\textrm{e}}=\frac{2\pi}{\hbar}\sum_{\mathbf q}\frac{|g_{\textrm{e},\mathbf q}|^2}{N}\delta(E_{c,\mathbf Q,\mathbf q}+\hbar\omega-E_{c,\mathbf K}), \nonumber\\
\frac{1}{\tau_\textrm{h}}=\frac{2\pi}{\hbar}\sum_{\mathbf q}\frac{|g_{\textrm{h},\mathbf q}|^2}{N}\delta(E_{v,\bm\Gamma,\mathbf q}-\hbar\omega-E_{v,\mathbf K'}).
\label{FermiGoldenRule}
\end{align}
Here, $\frac{1}{\sqrt{N}}g_{\textrm{e/h},\mathbf q}$ are the electron-phonon coupling matrix elements with $N$ the lattice number. In monolayer MoS$_2$, DFT calculation gives $g^0_{\textrm{e},\mathbf q}\sim0.11$ eV ($0.08$ eV) for LA ($A'_1$) phonons with wave vectors in the vicinity of $\mathbf M$, \cite{electron_phonon_coupling} and $g^0_{\textrm{h},\mathbf q}\sim0.1$ eV for TA phonons with wave vectors in the vicinity of $-\mathbf K'$.\cite{hole_phonon_coupling} From the interlayer coupling strength and band offset values given in Fig. \ref{Fig7}(c), we assume $20\%$ ($50\%$) of the involved $\mathbf Q_\textrm{c}$ ($\bm\Gamma_\textrm{v}$) valley in the heterobilayer is in the layer of the initial $\mathbf K$ electron ($\mathbf K'$ hole), which then leads to $g_{\textrm{e},\mathbf q}\sim\sqrt{0.2}g^0_{\textrm{e},\mathbf q}$ ($g_{\textrm{h},\mathbf q}\sim\sqrt{0.5}g^0_{\textrm{h},\mathbf q}$). We also use the effective mass approximation for the band dispersions $E_{c,\mathbf Q,\mathbf q}\approx E_{c,\mathbf Q}+\frac{\hbar^2q^2}{2m^*_{\mathbf Q}}$ and $E_{v,\bm\Gamma,\mathbf q}\approx E_{v,\bm\Gamma}-\frac{\hbar^2q^2}{2m^*_{\bm\Gamma}}$. Using the value $m^*_{\mathbf Q}\sim m_0$, \cite{Review_kpTheory} and taking into account both the LA, $A'_1$ phonons and the three possible pathways shown in Fig. \ref{Fig8}(b), we get $\tau_\textrm{e}\sim50$ fs. For the hole we use $m^*_{\bm\Gamma}\sim 2m_0$,\cite{Review_kpTheory}, which results in $\tau_\textrm{h}\sim50$ fs. They agree well with the experimental value ($<50$ fs) for the interlayer charge transfer process.\cite{Hong_NatNano,Yu_2015NL,Rigosi_2015NL}

The interlayer charge transfer mechanism proposed above is also consistent with the insensitivity to the interlayer twist, because the strong layer mixing nature of $\mathbf \Gamma_\textrm{v}$ and $\mathbf Q_c$ valleys is not affected. This is obvious for $\mathbf \Gamma_\textrm{v}$, where the interlayer coupling strength and band offset are not affected by the twist angle. For $\mathbf Q_c$ valleys, they are always on the ring region with strong interlayer coupling for any twist angle. Meanwhile, considering the large $\mathbf Q_c$ valley effective mass in the direction perpendicular to the $\mathbf \Gamma$-$\tau\mathbf K$ line,\cite{Review_kpTheory} the twist angle doesn't change the interlayer band offset much. Therefore the strong layer mixing of $\mathbf Q_c$ valleys are unaffected by the interlayer twist. For TMD heterobilayers with arbitrary stacking, the interlayer charge transfer can efficiently happen through emitting two intralayer phonons, one in the upper layer and the other in the lower layer.

\section{Conclusion}
In conclusion, the interlayer couplings in $\pm\mathbf K$, $\bm\Gamma_\textrm{v}$ and $\mathbf Q_\textrm{c}$ valleys of commensurate and incommensurate TMD bilayer structures are studied. The coupling strengths in $\pm\mathbf{K}$ valleys depend sensitively on the interlayer translation for R- and H-type commensurate bilayers, which can explain the observed band gap modulation in TMD heterobilayers with large scale moir\'{e} pattern. The coupling strengths for $\bm\Gamma_\textrm{v}$ and $\mathbf Q_\textrm{c}$ valleys are huge and insensitive to both the interlayer translation and twist angle. The resulted strong layer mixing of $\bm\Gamma_\textrm{v}$ and $\mathbf Q_\textrm{c}$ can mediate the twist-insensitive and ultrafast interlayer charge transfer in TMD heterobilayers. We expect that the results presented in this paper would be meaningful and illuminating for further exploring the rich physics and potential applications in various commensurate and incommensurate TMD bilayer structures.

\begin{acknowledgements}
We thank T. Cao for helpful discussions, and P. Rivera for proofreading. Y.W. and Z.W. were supported by NSFC with Grant No. 11604162 and Grant No.61674083. G.B.L. was supported by NSFC with Grant No. 11304014 and the China 973 Program with Grant No. 2013CB934500. H.Y. and W.Y. were supported by the Croucher Foundation (Croucher Innovation Award), the RGC and UGC of Hong Kong (HKU17305914P, AoE/P-04/08), and the HKU ORA.
\end{acknowledgements}

\end{document}